\documentclass[11pt]{article}
	
	\newcommand{\blind}{0}
	
    \makeatletter
    \renewcommand\section{\@startsection {section}{1}{\z@}%
                                       {-3.5ex \@plus -1ex \@minus -.2ex}%
                                       {2.3ex \@plus.2ex}%
                                       {\normalfont\fontfamily{phv}\fontsize{16}{19}\bfseries}}
    \renewcommand\subsection{\@startsection{subsection}{2}{\z@}%
                                         {-3.25ex\@plus -1ex \@minus -.2ex}%
                                         {1.5ex \@plus .2ex}%
                                         {\normalfont\fontfamily{phv}\fontsize{14}{17}\bfseries}}
    \renewcommand\subsubsection{\@startsection{subsubsection}{3}{\z@}%
                                        {-3.25ex\@plus -1ex \@minus -.2ex}%
                                         {1.5ex \@plus .2ex}%
                                         {\normalfont\normalsize\fontfamily{phv}\fontsize{14}{17}\selectfont}}
    \makeatother
	
	\usepackage{amsmath}
	\usepackage{graphicx}
	\usepackage{enumerate}
	\usepackage{amsfonts}
	\usepackage{amssymb}
	\usepackage{xcolor}
    \usepackage{algpseudocode}
    \usepackage[ruled]{algorithm2e}
    \usepackage{amsmath}
    \usepackage{mathtools}
	\usepackage{natbib} 
	\usepackage{url} 
    \usepackage{etoolbox}
    \usepackage[]{nomencl}   
    \makenomenclature
    
\newtheorem{remark}{Remark}

    \usepackage{xspace}
    \usepackage[compact]{titlesec}

	\DeclareMathOperator*{\argmaxB}{argmax} 

	\usepackage{hyperref} 
\hypersetup{
	colorlinks=true,       
	linkcolor=red,        
	citecolor=blue,        
	filecolor=magenta,     
	urlcolor=black
}

\begin{document}
		
		\def\spacingset#1{\renewcommand{\baselinestretch}%
			{#1}\small\normalsize} \spacingset{1}
		
		\if0\blind
		{ 		\title{ \bf A Novel Sparse Bayesian Learning and Its Application to Fault Diagnosis for Multistation Assembly Systems} 
		\author{Jihoon~Chung\textsuperscript{a}, Bo~Shen\textsuperscript{b}, and Zhenyu (James) Kong\textsuperscript{a} \\
		\textsuperscript{a}Grado Department of Industrial and Systems Engineering, Virginia Tech, Blacksburg, US\\ \textsuperscript{b}Department of Mechanical and Industrial Engineering,\\ New Jersey Institute of Technology, Newark, US }
			\date{}
			\maketitle
		} \fi
		
		\if1\blind
		{

            \title{\bf  Sparse Bayesian Learning with Temporally Correlated  Solution Vectors and Prior Knowledge with Application to Fault Diagnosis in Multistation Assembly Systems}
			\author{Author information is purposely removed for double-blind review}
			
\bigskip
			\bigskip
			\bigskip
			\begin{center}
				{\LARGE\bf \emph{IISE Transactions} \LaTeX \ Template}
			\end{center}
			\medskip
		} \fi
		\bigskip
		
\vspace{-0.4in}		\begin{abstract}
This paper addresses the problem of fault diagnosis in multistation assembly systems. Fault diagnosis is to identify process faults that cause the excessive dimensional variation of the product using dimensional measurements. For such problems, the challenge is solving an underdetermined system caused by a common phenomenon in practice; namely, the number of measurements is less than that of the process errors. To address this challenge, this paper attempts to solve the following two problems: (1) how to utilize the \textcolor{black}{temporal correlation in the time series data of each process error} and (2) how to apply prior knowledge regarding which process errors are more likely to be process faults. A novel sparse Bayesian learning method is proposed to achieve the above objectives. The method consists of three hierarchical layers. The first layer has parameterized prior distribution that exploits the temporal correlation of each process error. Furthermore, the second and third layers achieve the prior distribution representing the prior knowledge of process faults. Then, these prior distributions are updated with the likelihood function of the measurement samples from the process, resulting in the accurate posterior distribution of process faults from an underdetermined system. Since posterior distributions of process faults are intractable, this paper derives approximate posterior distributions via Variational Bayes inference. Numerical and simulation case studies using an actual autobody assembly process are performed to demonstrate the effectiveness of the proposed method. 
	\end{abstract}
			
	\noindent%
 	\setlength{\nomlabelwidth}{3cm}
    \nomenclature{$\lVert \textbf{x} \rVert_{p} $}{$\ell_{p}$ norm of the vector \textbf{x}.}%
    \nomenclature{$\lVert \text{A} \rVert_{F} $}{Frobenius norm of the matrix \text{A}.}%
    \nomenclature{$ \textbf{I}_{L}$}{Identity matrix with the size $L \times L$.}%
    \nomenclature{$ \text{A} \otimes \text{B}$}{Kronecker product of the matrices \text{A} and \text{B}.}%
    \nomenclature{$ \text{Vec(A)}$}{Vectorization of the matrix \text{A}.}%
    \nomenclature{$ \text{Tr(A)}$}{Trace of the matrix \text{A}.}%
    \nomenclature{$ \text{diag}\{ \alpha_{1},...,\alpha_{m} \}$}{Diagnoal matrix with principal diagonal elements being $\alpha_{1},...,\alpha_{m}$.}%
     \printnomenclature	
	
	\spacingset{1.5} 

\section{Introduction} \label{s:intro}
Multistation assemblies represent the systems that perform assembly operations on multiple stations to assemble a final product. The quality of the final product relies on several factors, called key control characteristics (KCCs) or process errors \citep{bastani2018fault}. The positioning accuracy of fixture locators is KCC in the multistation assembly \citep{bastani2012fault}. Fixture locators clamp the parts during the assembly process, so their deviation from nominal may cause dimensional quality problems in the final product. Therefore, it is necessary to identify process errors that have mean shifts and/or large variance increases from their design specifications, namely, process faults \citep{bastani2016compressive}. Hence, fault diagnosis in multistation assemblies estimates the mean and variance of process errors, namely, the variations of fixture locators. This paper focuses on the process fault of the mean shift.

Because of the unduly cost and physical constraints, the dimensional variation of KCCs, namely, fixture locators, cannot be directly monitored using sensors in multistation assemblies \citep{lee2020variation}. Instead, key product characteristics (KPCs), namely, the key measurements from the final product, can be used to estimate the KCCs and, consequently, identify process faults among process errors. \textcolor{black}{A fault-quality linear model of multistation process} represents the relationship between KPCs and process errors as follows \citep{huang2007stream,huang2007stream1}:
\begin{equation}\label{eq:1}
    \textbf{y}=\Phi \textbf{x} +\textbf{v},
\end{equation}
where $\textbf{y}\in \mathbb{R}^{M \times 1} $represents $M$ dimensional measurements (i.e., KPCs), $ \textbf{x}\in \mathbb{R}^{N \times 1} $denotes $N$ process errors \textcolor{black}{(i.e., KCCs}), $\Phi \in \mathbb{R}^{M\times N}$ is a fault pattern matrix obtained \textcolor{black}{from all the process information in the multistage process \citep{ding2000modeling}}, and $\textbf{v} \in \mathbb{R}^{M \times 1}$ denotes the noise. \textcolor{black}{Although the relationship between KPCs and KCCs is nonlinear in many manufacturing processes, a fault-quality linear model is used in general as the nonlinear relationship could be approximated to a linear model utilizing Taylor series expansion because the variations of KCCs are small, and the relationship between KCCs and KPCs is smooth and without sharp changes \citep{lee2020variation, shi2022process,ding2002fault}.} Since process errors indicate the mean deviations of the fixture locators, process faults refer to nonzero elements in $\textbf{x}$, namely, process errors with nonzero mean shifts \citep{bastani2018fault}.  

Following the fault-quality model in Eq.~\eqref{eq:1}, some research on fault diagnosis for manufacturing systems has been investigated \citep{ding2002fault,zhou2004statistical}. All these methods assume that the number of measurements ($M$) is greater than the number of process errors ($N$) (i.e., $M>N$). However, this assumption may not always hold in actual manufacturing applications, as using an excessive number of sensors (measurements) will result in undue costs \citep{bastani2018fault}. These approaches are unsuitable if this assumption is violated. This is because Eq.~\eqref{eq:1} becomes an underdetermined system that results in the non-existence of a unique solution. To overcome the challenge, the sparse solution assumption \citep{donoho2006most} that \textbf{x} in Eq.~\eqref{eq:1} has a minimal number of nonzero elements is required. In the context of the fault diagnosis problem, it denotes the sparsity of process faults in the fault-quality linear model. This is reasonable since it is likely to have a few process faults in practice \citep{bastani2012fault}. 
Among the several sparse estimation methods, the Bayesian method called sparse Bayesian learning has received much attention recently because of its superior estimation performance \citep{zhang2011sparse}. 

Several studies have used sparse Bayesian learning for fault diagnosis in manufacturing systems \citep{lee2020variation,bastani2018fault,bastani2012fault,li2016bayesian}. These studies successfully identified process faults by providing prior distribution of process errors (i.e., \textbf{x} in Eq.~\eqref{eq:1}) to promote the sparsity of process faults. Especially, the work in \citep{bastani2018fault} applied Bayesian learning to diagnose mean shift fault, which is the most relevant work to our study.  In \citep{bastani2018fault}, \textcolor{black}{multiple KPCs samples} have been used to estimate the mean deviation of fixture locators. Specifically, given the average of \textcolor{black}{multiple KPCs samples}, which is $\textbf{y}$ in Eq.~\eqref{eq:1}, the mean deviation of the fixture locators (namely, $\textbf{x}$ in Eq.~\eqref{eq:1}) can be estimated. However, this work does not consider the characteristics of multiple KPCs samples where they are collected sequentially. \textcolor{black}{In practice, time series data of each process error from the multiple KPCs samples may have a strong temporal correlation in the multistation assembly process. For example, the locator position of the fixture system could have a drifting due to wear \citep{zhou2004statistical}, causing dimensional quality issues in the samples. If the process faults are not mitigated immediately, the effect of fixture deviation is auto-correlated in terms of time due to the degradation of wear of production tooling over time \citep{shi2022process}. Similarly, the dimensional variability caused by machine-tool thermal distortions in the multistation processes is highly correlated between the samples assembled in a certain period \citep{abellan2012state}. Therefore, the temporal correlation of each process error causes product samples manufactured over a period of time to exhibit the same patterns of faults by a specific source of variation \citep{liu2010variation}.} The relationship between the sequentially collected \textcolor{black}{multiple KPCs samples} and their process errors can be formulated as the following multiple measurements vectors (MMV) model \citep{cotter2005sparse} that extends from Eq.~\eqref{eq:1}:
\begin{equation}\label{eq:3}
      \textbf{Y}=\Phi \textbf{X} +\textbf{V},
\end{equation}
where $\textbf{Y}=[\textbf{Y}_{\cdot 1},...,\textbf{Y}_{\cdot L}] \in \mathbb{R}^{M\times L}$ is a measurement matrix consisting of \textcolor{black}{$L$  KPCs samples}, and $\textbf{Y}_{\cdot i} \in \mathbb{R}^{M \times 1}$ is a vector that denotes the \textcolor{black}{$i^{th}$ KPCs sample}. $\textbf{X}=[\textbf{X}_{\cdot 1},...,\textbf{X}_{\cdot L}] \in \mathbb{R}^{N\times L}$ is a process error matrix, where $\textbf{X}_{\cdot i} \in \mathbb{R}^{N \times 1}$ is a \textcolor{black}{vector that represents process errors (KCCs) of the $i^{th}$ KPCs sample.} $\textbf{V}\in\mathbb{R}^{M\times L}$ is a noise matrix. Since \textcolor{black}{$L$ KPCs samples} have the same process faults, all columns of $\textbf{X}$ share the index of nonzero rows called support. It is called a common support assumption in the MMV model \citep{cotter2005sparse}. \textcolor{black}{ In addition,  $L$ elements in the $j^{th}$ row of $\textbf{X}$, representing the time series data of the $j^{th}$ process error,  are highly correlated as nonzero values if the $j^{th}$ process error is process fault.} \textcolor{black}{ However, the dynamic changes of process faults due to the complexity of the manufacturing process can easily violate the common support assumption in a large number of KPCs samples. 
Therefore, this paper focuses on the small number of KPCs samples ($L$) to satisfy the common support assumption. Utilizing a small number of KPCs samples is also efficient in fault diagnosis of process faults in the multistation assembly process for time and cost reduction.}

\textcolor{black}{Beyond the common support assumption and temporal correlation in the time series data of each process error,} utilizing prior knowledge of process faults is an additional way to improve the identification of sparse process faults \citep{lee2020variation}. Specifically, the prior knowledge regarding which process errors are more likely to be process faults than others. This knowledge can be obtained from domain-specific knowledge from practitioners or collected based on the fault diagnosis at the past time stamps. For example, the manufacturing engineers in an assembly line usually know that some fixture locators may malfunction more frequently than others based on their experiences. In practice, the prior knowledge provides only part of the actual process faults. In addition, the knowledge may contain some erroneous information as to process faults, which are actually not. Therefore, the prior knowledge provides partial and even erroneous information about process faults. However, utilizing the prior knowledge is expected to improve the identification of process faults if the correct and erroneous information can be properly distinguished. 

In the sparse Bayesian learning  literature, there exist studies that consider the temporal correlation  \textcolor{black}{of rows in matrix \textbf{X}} in Eq.~\eqref{eq:3} based on the common support assumption \citep{luessi2013bayesian,han2018bayesian}, and the work that utilizes the partial with some erroneous prior knowledge of support to improve the performance of sparse estimation \citep{fang2015support,guo2017variational}, separately. However, these studies did not integrate these aspects to improve the sparse estimation.

To address this challenge, this paper aims to develop a novel sparse Bayesian hierarchical learning method that simultaneously utilizes the \textcolor{black}{temporal correlation in the time series data of each process error}, as well as prior knowledge, which may contain erroneous information. \textcolor{black}{The new method, namely, support knowledge aided with temporally correlated process error sparse Bayesian learning (SA-TSBL)}, is proposed to achieve this objective. The contributions of this work are summarized as follows: 
\vspace{-0.3cm}\begin{itemize}
    	\item From the methodological point of view, this paper proposes a novel sparse Bayesian learning that considers both \textcolor{black}{temporal correlation in the time series data of each process error and prior knowledge of process faults} to improve the sparse estimation. This method also derives an approximate posterior distribution of the sparse solution via Variational Bayes inference \citep{petersen2005slow} to address the intractable computational challenge.\vspace{-0.3cm}
    	\item 	From the application perspective, the proposed method is applied to fault diagnosis in the multistation assembly systems. The method mitigates the dimensional quality issues in the assembly operation by effectively identifying the fixture locators with excessive mean shifts. The effectiveness of the proposed method is validated in real-world simulation case studies that use an actual auto body assembly process. 
\end{itemize}\vspace{-0.3cm}

The rest of this paper is organized as follows. A brief review of related research work is provided in Section~\ref{s:sec2}. The proposed methodology is presented in Section~\ref{s:sec3}, followed by numerical case studies to validate its effectiveness in Section~\ref{s:sec4}. Section~\ref{s:sec5} offers real-world case studies on fault diagnosis problems in the multistation assembly process. Finally, conclusions and future work are discussed in Section~\ref{s:sec6}.

\section{Review of Related Work} \label{s:sec2}
The related existing studies of fault diagnosis in manufacturing systems are reviewed in Section~\ref{s:sec2.1}. Then, the literature related to sparse Bayesian learning is provided in Section~\ref{s:sec2.2}. Afterward, the research gaps in the current work are identified in Section~\ref{s:sec2.3}. 
\subsection{\emph{Fault Diagnosis Methodologies in Manufacturing Systems}} \label{s:sec2.1}
There has been a large body of research on fault diagnosis methodologies for manufacturing systems based on the fault-quality model in Eq.~\eqref{eq:1}. \cite{kong2008multiple} developed a PCA-based orthogonal diagonalization strategy to transform the measurement data. It enabled the estimation of the variance of process errors in a multistation assembly system. \cite{zhou2004statistical} applied a mixed linear model to represent the relationship between measurements and process faults in a multistage machine process. Then, it used a maximum likelihood estimator for both mean shift and variance change to detect the process change. \cite{ceglarek1996fixture} presented a fault diagnosis method based on fault mapping procedures that combine principal component analysis (PCA) and pattern recognition.
All these approaches assume the number of measurements is greater than the number of the process errors (i.e., $M>N$ in Eq.~\eqref{eq:1}), which may not be consistent with the industrial practice. However, if this assumption is violated, all of the approaches mentioned above are ineffective because the fault-quality linear model becomes an underdetermined system resulting in the non-existence of a unique solution. 

\textcolor{black}{
To overcome an underdetermined system for fault diagnosis in the manufacturing system, sparse learning which has been actively used in fault diagnosis and detection in the manufacturing system can be considered. \cite{han2018intelligent} developed a fault diagnosis method via dictionary learning and sparse representation-based classification for fault diagnosis in the manufacturing process. 
\cite{zhang2022bearing} proposed a new non-convex penalty called the generalized logarithm penalty, which enables sparsity and reduces noise disturbance for bearing fault diagnosis. 
\cite{zhang2021dynamic} developed a dynamical multivariate functional data modeling via sparse subspace learning to detect process faults in the manufacturing processes. \cite{dai2021group} proposed a group-sparsity learning approach for bearing fault diagnosis. In addition to fault diagnosis and detection in the general manufacturing process, sparse learning has been widely utilized to address the issue of an underdetermined system in the multistage assembly system. Specifically, sparse Bayesian learning has been actively utilized to incorporate the sparsity of process faults as the prior distribution.} \cite{bastani2012fault} proposed a fault diagnosis approach by combining the state-space model and the relevant vector machine to figure out process faults using the sparse estimate of the variance change of process errors. \cite{li2016bayesian} developed a Bayesian variable selection-based method to identify both process faults and sensor faults in the assembly process. \cite{bastani2018fault} proposed a spatially correlated sparse Bayesian learning to deal with the case when process errors have a spatial correlation. The work is based on the hypothesis that if one of the fixture locators deviates from its design specification, the neighboring locators are also expected to deviate. \cite{lee2020variation} presented a Bayesian approach for identifying variation sources in a multistage manufacturing process using the sparse variance component prior. The work focuses on the identification of process faults that have variance increases.
\subsection{\emph{Sparse Bayesian Learning }} \label{s:sec2.2}
Since \cite{tipping2001sparse} proposed Sparse Bayesian learning (SBL), many researchers have significantly extended it. \cite{wipf2009solving} first introduced SBL to sparse estimation for the single measurement vector model in Eq.~\eqref{eq:1}. Then \cite{wipf2007empirical} extended it to the MMV model (Eq.~\eqref{eq:3}), deriving the MSBL algorithm using the common support assumption. \textcolor{black}{The advantage of SBL and MSBL is that global minimum of both methods are always the sparsest solution compared to the minimization based algorithms \citep{tibshirani1996regression, chen2001atomic}, where global minimum is usually not the sparsest solution \citep{zhang2011sparse, candes2008enhancing}. In addition, sparse Bayesian learning have much fewer local minima than sparse learning with frequentist approaches such as the FOCUSS family \citep{zhang2011sparse, cotter2005sparse}.} Based on the MMV model, several studies exploited \textcolor{black}{the temporal correlation} to improve the performance of sparse estimation. \cite{zhang2011sparse} presented a block SBL framework where a positive definite matrix captures the correlation structure of each row of \textbf{X} in Eq.~\eqref{eq:3}. \cite{luessi2013bayesian} established a hierarchical Bayesian framework to model the temporally smooth signals using a multinomial distribution as the prior distribution of each row of \textbf{X} in Eq.~\eqref{eq:3}. \citep{han2018bayesian} provided a Wishart distribution as the prior distribution to learn the temporal correlation from multiple measurement samples. Besides considering the \textcolor{black}{temporal correlation}, work utilizing prior knowledge of the support has been actively studied recently under the SBL framework. To utilize the prior knowledge in the Bayesian framework, \cite{fang2015support,li2017adaptive, yu2019sa} added one more hierarchical layer to conventional SBL to integrate the prior knowledge of support as the prior distribution. 
\subsection{
\emph{Research Gap Analysis }} \label{s:sec2.3}
The work summarized in Section~\ref{s:sec2.1} investigates process faults in multistation assembly systems for quality assurance. Recently, the sparsity of process faults has been used via SBL to deal with the low dimensional measurements in fault diagnosis problems, which is common in industrial practice. However, there is a lack of efforts to identify process faults by considering \textcolor{black}{temporal correlation in the time series data of each KCC} and utilizing prior knowledge of process faults. Research efforts in Section~\ref{s:sec2.2} introduce methods in SBL that consider the temporal correlation and prior knowledge of support. However, these two aspects were not integrated into the reported work. Therefore, this paper proposes a novel SBL method that simultaneously utilizes \textcolor{black}{temporal correlation in the time series data of each KCC} and prior knowledge of process faults to improve the identification of sparse process faults in multistation systems.
\section{Proposed Research Methodology} \label{s:sec3}
This section proposes a novel sparse Bayesian hierarchical method: support knowledge aided \textcolor{black}{temporally correlated process error SBL (SA-TSBL).} The proposed \textcolor{black}{SA-TSBL} is described in Section~\ref{s:methods.3.1}, followed by Bayesian inference of the proposed method in Section~\ref{s:methods.3.2}.
\subsection{\emph{Proposed Methodology}} \label{s:methods.3.1}
\begin{figure}[!ht]
    \centering
    \resizebox*{11cm}{!}{\includegraphics{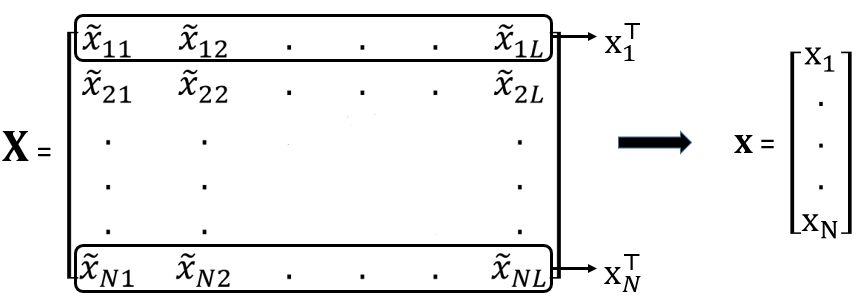}}\vspace{-0.5cm}
    \caption{Block structure of \textbf{x}.}
    \label{fig:fig1}
\end{figure}

The proposed methodology is a sparse Bayesian hierarchical model using \textcolor{black}{multiple KPCs samples} that have the same process faults. The method considers the \textcolor{black}{correlation in the time series data of each process error,} and utilizes prior knowledge of process faults to improve the sparse estimation. To exploit the \textcolor{black}{temporal correlation in the time series data of each process error}, the proposed method transforms the MMV model in Eq.~\eqref{eq:3} to the following block single measurement vector model \citep{zhang2011sparse}. 
\begin{equation}\label{eq:4}
    \textbf{y}=\textbf{D} \textbf{x} +\textbf{v},
\end{equation}
where \textcolor{black}{$\textbf{y}=$Vec$(\textbf{Y}^{\top})\in\mathbb{R}^{ML\times 1}, \textbf{D}=\Phi \otimes \textbf{I}_{L}, \textbf{x}=$Vec$(\textbf{X}^{\top})\in \mathbb{R}^{NL\times1}$. Assume noise vector \textbf{v} follows Gaussian distribution with zero mean and variance $\lambda$.} Eq.~\eqref{eq:4} can be rewritten as
\begin{equation}\label{eq:5}
    \textbf{y}= [\phi_{1}\otimes\textbf{I}_{L},...,\phi_{N}\otimes\textbf{I}_{L}][\text{x}_{1}^{\top},...,\text{x}_{N}^{\top}]^{\top}+\textbf{v},
\end{equation}
where $\phi_{i}$ is the $i^{th}$ column in \textbf{$\Phi$}, and $\text{x}_{i}$ consists of the $i^{th}$ process errors of $L$ KPCs samples (i.e., $\text{x}_{i}^{\top}=(\tilde{x}_{i1}, \tilde{x}_{i2},...,\tilde{x}_{iL})$, where $\tilde{x}_{ij}$ denotes the element in the $i^{th}$ row and the $j^{th}$ column of matrix $\textbf{X}$ in Eq.~\eqref{eq:3}). In other words, $\text{x}_{i}\in \mathbb{R}^{L \times 1}$ is the $i^{th}$ block of \textbf{x} in Eq.~\eqref{eq:4}, as illustrated in Figure~\ref{fig:fig1}. 
\vspace{-0.3cm}\vspace{-0.3cm}
Therefore, $K$ nonzero rows in \textbf{X} in Eq.~\eqref{eq:3} are represented as $K$ nonzero blocks in \textbf{x} in the proposed method. It implies that there exist $K$ process errors that shifted from the design specification, namely, process faults. \textcolor{black}{Since $\text{x}_{i}$ consists of the time series data of $i^{th}$ process error from $L$ KPCs samples}, the proposed method exploits the correlation between the $L$ elements of $\text{x}_{i}$. \textcolor{black}{Since noise vector \textbf{v} follows Gaussian distribution with zero mean and variance $\lambda$, the Gaussian likelihood is provided for the block model in Eq.~\eqref{eq:4} as follows:} 
\begin{equation*}
    p(\textbf{y}|\textbf{x};\lambda)\sim N(\textbf{Dx},\lambda\textbf{I}_{ML}).
\end{equation*}

\begin{figure}[!ht]
    \centering
    \resizebox{0.9\textwidth}{!}{\includegraphics{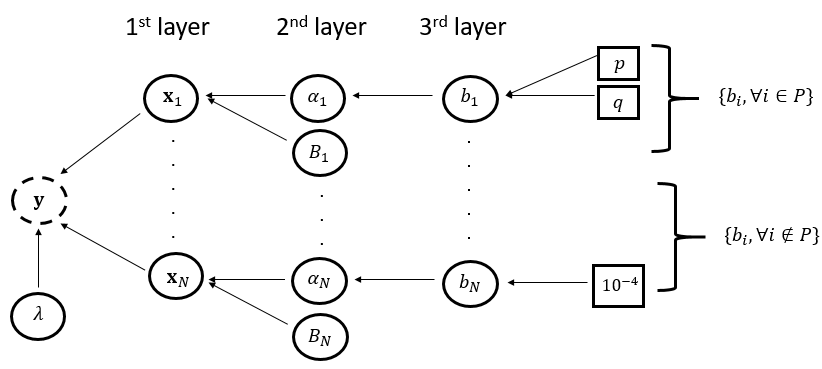}}\vspace{-0.5cm} 
    \caption{Graphical representation of the proposed method. A circle indicates a random variable or hyperparameter that needs to be estimated. A dashed circle and square represent an observation and a constant, respectively.}
    \label{fig:fig2}
\end{figure}

The proposed method consists of the following three layers. The prior distribution in the first layer is provided to exploit the \textcolor{black}{correlation of the time series data of each KCC}. The second and third layers consist of prior distribution representing the prior knowledge of process faults among process errors. A graphical representation of the proposed method is shown in Figure~\ref{fig:fig2}. 
\vspace{-0.0cm}\vspace{-0.0cm}

In the first layer of the hierarchical model in Figure~\ref{fig:fig2}, \textcolor{black}{the Gaussian distribution is provided as the prior distribution for process error (\textbf{x}) since it is the most common distribution considered for process errors in the literature and practice \citep{lee2020variation, bastani2012fault}. The prior distribution is given by}
\begin{equation*}
    p(\textbf{x}|\boldsymbol{\alpha};\textbf{B}) \sim N(\textbf{0},\Sigma_{0}),
\end{equation*}
where $\Sigma_{0}$ is 
$$
\Sigma_{0}=\left[
\begin{array}{ccc}
   \alpha_{1}^{-1}\text{B}_{1}^{-1}  & \cdots & 0 \\
   \vdots & \ddots & \vdots \\
   0 & \cdots & \alpha_{N}^{-1}\text{B}_{N}^{-1}
\end{array}
\right],
$$
$\boldsymbol{\alpha}=\{ \alpha_{1},...,\alpha_{N}\}$, and $\textbf{B}=\{ \text{B}_1,...,\text{B}_N \}$. Since $\text{x}_{i}$ consists of the \textcolor{black}{time series data of $i^{th}$ process error from $L$ KPCs samples}, $\alpha_{i}^{-1}$ controls the sparsity of the $i^{th}$ process error ($\text{x}_{i}$). For example, when $\alpha_{i}^{-1}$ converges to zero, the associated block $\text{x}_{i}$ will be driven to zero \citep{tipping2001sparse}. $\text{B}_{i}$ is a positive definite matrix that \textcolor{black}{captures the correlation of the time series data of $i^{th}$ process error. The proposed method assumes the independence between process errors (e.g., different fixture locators), which is a common assumption in the literature \citep{lee2020variation, bastani2016compressive, li2016bayesian}. It results in the block diagonal matrix $\Sigma_{0}$.} 

The second layer in Figure~\ref{fig:fig2} specifies Gamma prior distributions over $\boldsymbol{\alpha}$. The distributions have an individual rate parameter $b_{i}$ for each parameter $\alpha_{i}$, namely,
\begin{equation}\label{eq:6}
    p(\boldsymbol{\alpha}|\boldsymbol{b})=\prod_{i=1}^{N}\text{Gamma}(\alpha_{i}|a,b_{i})=\prod_{i=1}^{N}\Gamma(a)^{-1}b_{i}^{a}\alpha_{i}^{a-1}e^{-b_{i}\alpha_{i}},
\end{equation}
where $\boldsymbol{b}=\{b_{1},...,b_{N}\}$. Eq.~\eqref{eq:6} effectively integrates the prior information of process faults into the SBL framework. Basically, $a$ and $b_{i}$ are set to be very small values (e.g., $10^{-4}$) to provide a large variance of a prior distribution over $\alpha_{i}$ in SBL \citep{tipping2001sparse}. It encourages the large values of $\alpha_{i}$, and promotes the $i^{th}$ process error as non-process faults (i.e., $\text{x}_{i}=\boldsymbol{0}$ in Eq.~\eqref{eq:5}) \citep{fang2015support}. \textcolor{black}{Specifically, the marginal distribution of the time series data of $i^{th}$ process error, namely, $p(\text{x}_{i})=\int p(\text{x}_{i}|\alpha_{i}, \text{B})p(\alpha_{i}|a, b_{i})\text{d}\alpha_{i}$, follows multivariate student-t distribution which the probability is concentrated at zero \citep{tipping2001sparse}.} Suppose set $P$ consists of the indexes of process errors that are likely to be process faults based on prior knowledge. Then, the corresponding rate parameters $\{b_{i}, i\in P \}$ are set as a relatively large value (e.g., 1) to provide a small variance of the prior distribution over $\alpha_{i}$ \citep{fang2015support}. The prior distribution allows the small value of $\alpha_{i}$, and encourages the $i^{th}$ process error as process faults (i.e., $\text{x}_{i}\neq 0$ in Eq.~\eqref{eq:5}). However, assigning a fixed value to $b_{i}$ in Eq.~\eqref{eq:6} has limitations to deal with the situation when the set $P$ has some erroneous information of process faults. 

To address this issue, the third layer in Figure~\ref{fig:fig2} assigns a prior distribution over $\{b_{i}, i\in P \}$ \citep{li2017adaptive}. \textcolor{black}{Gamma distribution is assigned because of the support of the variable. In addition, the distribution enables users to derive the closed form of the approximate posterior distribution of $\{b_{i}, i\in P \}$, which will be described in the following section. The prior distribution over $\{b_{i}, i\in P \}$ is provided as follows:}
\begin{equation}\label{eq:7}
   \text{Gamma}(b_{i}|p,q)=\Gamma(p)^{-1}q^{p}b_{i}^{p-1}e^{-qb_{i}}, i\in P.
\end{equation}
$p$ and $q$ in Eq.~\eqref{eq:7} are specified to characterize prior distribution. These two values in the proposed method can be adjusted by the users to satisfy the following two conditions as follows. 
\vspace{-0.3cm}\begin{itemize}
\item First, the values are set to provide a large mean value of the prior distribution of $\{b_{i}, i\in P \}$. \vspace{-0.3cm}
\item Second, the values are determined to have a large variance in the prior distribution of $\{b_{i}, i\in P \}$. 
\end{itemize}\vspace{-0.3cm}

The first condition encourages the $i^{th}$ process error as process faults based on prior knowledge. Compared to a small fixed value of $b_{i}=10^{-4}$ that promotes the large $\alpha_{i}$ in previous research \citep{fang2015support,tipping2001sparse}, the large mean value of the prior distribution of $\{b_{i}, i\in P \}$ lets the small $\alpha_{i}$. The small $\alpha_{i}$ provides a prior distribution that $i^{th}$ process error is likely to be process faults of the mean shift.

The second condition is designed to deal with the case when erroneous prior knowledge is provided. Assume incorrect prior knowledge that the $i^{th}$ process error may be a process fault is provided. If the variance of the prior distribution of $\{b_{i}, i\in P \}$ is small, the $i^{th}$ process error is likely to be misdiagnosed as process faults because of the large mean value of the prior distribution of $\{b_{i}, i\in P \}$ from the first condition. However, if the variance of the prior distribution of $\{b_{i}, i\in P \}$ is large, $\{b_{i}, i\in P \}$ is not significantly affected by the prior distribution. Instead, $\{b_{i}, i\in P \}$ is highly affected by the data itself, where the $i^{th}$ process error is not the process fault. This allows $\{b_{i}, i\in P \}$ to be learned as a small value from data and provides the sparsity prior to the $i^{th}$ process error \citep{fang2015support}.

For $\{b_{i}, i\in P^{c}\}$ in the proposed method, the parameters are set to a fixed small value of $10^{-4}$ as in the previous study \citep{tipping2001sparse}.

\subsection{\emph{Bayesian Inference of the Proposed Methodology}} \label{s:methods.3.2}
The proposed method in Section~\ref{s:methods.3.1} has several hidden variables that need to be estimated, that is, the process errors (\textbf{x}), the variable controlling the sparsity of process errors ($\boldsymbol{\alpha}$), and the variable used to provide prior knowledge of process faults ($b_{i}$, $\forall i\in P$). In addition, there are hyperparameters related to \textcolor{black}{temporal correlations in the time series data of $i^{th}$ KCCs ($\text{B}_{i}$) and noise ($\lambda$),} which also require estimation. \textcolor{black}{To avoid too many parameters to being estimated, causing a challenging task in sparse estimation, the proposed method sets $\text{B}_{i}=\text{B}$ $(\forall i)$ \citep{zhang2011sparse}.}  

\textcolor{black}{However, the posterior distribution of hidden variables in the proposed method (Eq.~\eqref{eq:80}) does not have a closed form because of the complexity of the proposed hierarchical model in Section~\ref{s:methods.3.1}. Specifically, the denominator in Eq.~\eqref{eq:80} cannot be calculated as closed form. Let $b_{i}$ $(i\in P)$ as $\bar{b}$ for convenience. 
\begin{align}\label{eq:80}
         P(\textbf{x}, \boldsymbol{\alpha}, \bar{b} | \textbf{y}) & = \frac{P(\textbf{y}, \textbf{x}, \boldsymbol{\alpha},\bar{b})}{  \idotsint\limits_{\textbf{x}, \boldsymbol{\alpha}, \bar{b} } P(\textbf{y}, \textbf{x}, \boldsymbol{\alpha},\bar{b}) \, d\textbf{x} \, d\boldsymbol{\alpha} \, d\bar{b} }      
\end{align}}\vspace{-0.6cm}

\textcolor{black}{To overcome this challenge, this paper derives approximate posterior distributions of hidden variables via Variational Bayes inference (VBI). Specifically, Variational Bayes Expectation Maximization (VBEM) \citep{petersen2005slow} is utilized to estimate hidden variables and hyperparameters in the proposed method to identify mean shifts process faults. VBEM consists of (1) E-step: Variational Bayesian expectation step to estimate hidden variables $\textbf{x}, \boldsymbol{\alpha}, \bar{b}$ by approximating the posterior distribution of hidden variables; and (2) M-step: Variational Bayesian maximization step to update hyperparameters $\text{B}$ and $\lambda$ by maximizing the expected value of the logarithm of the complete likelihood.}

Let $\boldsymbol{\theta}$ be a vector with all hidden variables in the proposed method (i.e., $\boldsymbol{\theta}= (\textbf{x}, \boldsymbol{\alpha} ,\bar{b}))$. VBI approximates the posterior distribution of $\boldsymbol{\theta}$, denoted as $q(\boldsymbol{\theta})$, by minimizing Kullback-Leibler (KL) divergence between $q(\boldsymbol{\theta})$ and the true posterior distribution, namely, $p(\boldsymbol{\theta}|\boldsymbol{y})$ (i.e., $D_{KL} (q(\boldsymbol{\theta})||p(\boldsymbol{\theta}|\boldsymbol{y})$)). $q(\boldsymbol{\theta})$ is factorized as 
\begin{equation*}
    q(\boldsymbol{\theta})=q(\textbf{x})q(\boldsymbol{\alpha})q(\bar{b})
\end{equation*}
by the mean-field approximation \citep{cohn2010mean}. \textcolor{black}{The approximate posterior distribution $q(\theta_{i})$, where $\theta_{i}$ is the $i^{th}$ element in the set $\boldsymbol{\theta}$ is derived as follows by minimizing the $D_{KL} (q(\boldsymbol{\theta})||p(\boldsymbol{\theta}|\boldsymbol{y})$ under the mean-field approximation. 
\begin{equation}\label{eq:8}
\ln{q(\theta_{i})} = \mathbb{E}[\ln{p(\boldsymbol{y},\boldsymbol{\theta}})]_{\boldsymbol{\theta} \setminus \theta_{i}}+const,
\end{equation}
where $\mathbb{E}_{\boldsymbol{\theta} \setminus \theta_{i}}$ denotes the expectation taken with the set $\boldsymbol{\theta}$ without $\theta_{i}$.} const can be obtained through normalization. Eq.~\eqref{eq:8} is used in the following E-step of VBEM to approximate the posterior distributions of hidden variables.

\textbf{E-step of VBEM}: The posterior distributions of hidden variables that are related to process errors ($\textbf{x}$), sparsity of process errors ($\boldsymbol{\alpha}$), and prior knowledge of process faults ($\bar{b}$) are approximated by Eq.~\eqref{eq:8}, respectively, as follows.
\begin{align}\label{eq:9}
      \ln{q(\textbf{x})} &= \langle \ln{p(\textbf{y},\textbf{x}, \boldsymbol{\alpha}, \bar{b})} \rangle_{q(\boldsymbol{\alpha})q(\bar{b})}+const  \nonumber \\
      & =\langle \ln{p(\textbf{y}|\textbf{x};\lambda})p(\textbf{x}|\boldsymbol{\alpha};\textbf{B}) \rangle_{q(\boldsymbol{\alpha})}+const,
\end{align}
\begin{align}\label{eq:10}
      \ln{q(\boldsymbol{\alpha})} &= \langle \ln{p(\textbf{y},\textbf{x}, \boldsymbol{\alpha}, \bar{b})} \rangle_{q(\textbf{x})q(\bar{b})}+const  \nonumber \\
      & =\langle \ln{p(\textbf{x}|\boldsymbol{\alpha};\textbf{B}})p(\boldsymbol{\alpha}|\bar{b}) \rangle_{q(\boldsymbol{\alpha})}+const,
\end{align}
\begin{align}\label{eq:11}
      \ln{q(\bar{b})} &= \langle \ln{p(\textbf{y},\textbf{x}, \boldsymbol{\alpha}, \bar{b})} \rangle_{q(\textbf{x})q(\boldsymbol{\alpha})}+const  \nonumber \\
      & =\langle \ln{p(\boldsymbol{\alpha}|\bar{b}})p(\bar{b}) \rangle_{q(\boldsymbol{\alpha})}+const,
\end{align}
where $\langle \cdot \rangle$ indicates the expectation. Based on the statistical inference, the posterior distributions of hidden variables can be derived as
\begin{equation}\label{eq:12}
    q(\textbf{x})=N(\textbf{x}|\mu_{\textbf{x}},\Sigma_{\textbf{x}}),
\end{equation}
\begin{equation}\label{eq:13}
    q(\boldsymbol{\alpha})=\prod_{i=1}^{N}\text{Gamma}(\alpha_{i}|\tilde{a},\tilde{b}_{i}),
\end{equation}
\begin{equation}\label{eq:14}
    q(\bar{b})=\prod_{i\in P}\text{Gamma}(b_{i}|p+a,\langle \alpha_{i} \rangle+q).
\end{equation}
The expectations and moments of distributions in Eqs. \eqref{eq:12}, \eqref{eq:13}, and \eqref{eq:14} are
\begin{equation}\label{eq:15}
    \mu_{\textbf{x}} = \frac{1}{\lambda}\Sigma_{\textbf{x}}\textbf{D}^{\top}\textbf{y},
\end{equation}
\begin{equation}\label{eq:16}
    \langle \alpha_{i} \rangle = \frac{a+\frac{L}{2}}{\langle {b}_{i} \rangle}, \hspace{0.25em} (\forall i),
\end{equation}
\begin{equation}\label{eq:17}
    \langle b_{i} \rangle = \frac{p+a}{q+\langle a_{i}\rangle },  \hspace{0.25em}(i\in P),
\end{equation}
\begin{equation}\label{eq:18}
    \Sigma_{\textbf{x}} = (\frac{1}{\lambda}\textbf{D}^{\top}\textbf{D}+\langle \text{AB} \rangle)^{-1},
\end{equation}
where $\langle \text{AB} \rangle = \text{diag}[\langle \alpha_{1} \rangle \text{B},...,\langle \alpha_{N} \rangle\text{B}]$. \textcolor{black}{The estimator of the time series data of $i^{th}$ process error} ($\langle\text{x}_{i}\rangle$) can be obtained from $\mu_{\textbf{x}}$ in Eq.~\eqref{eq:15} as follows.
\begin{equation}\label{eq:100}
\langle \text{x}_{i} \rangle = \mu_{\textbf{x}}((i-1)L+1:iL), i=1,...,N,
\end{equation}
Detailed derivations of Eqs.~\eqref{eq:12},~\eqref{eq:13}, and~\eqref{eq:14} are provided in the Appendices~\ref{app:app1},~\ref{app:app2}, and~\ref{app:app3}, respectively.

\textbf{M-step of VBEM}: \textcolor{black}{Temporal correlations in the time series data of each KCC (\text{B}) and noise ($\lambda$)} are estimated in this step. Let $\tilde{\boldsymbol{\theta}}=\{ \text{B},\lambda \}$. Posterior distributions of $\textbf{x}, \boldsymbol{\alpha} ,\bar{b}$ obtained in Eqs. \eqref{eq:12}, \eqref{eq:13}, and \eqref{eq:14} are denoted as $q(\textbf{x};\tilde{\boldsymbol{\theta}}^{OLD}), q(\boldsymbol{\alpha};\tilde{\boldsymbol{\theta}}^{OLD})$, and $q(\bar{b};\tilde{\boldsymbol{\theta}}^{OLD})$ respectively. $\tilde{\boldsymbol{\theta}}$ can be updated by maximizing the complete likelihood as follows:
\begin{equation*}
\begin{split}
        \tilde{\boldsymbol{\theta}}^{NEW} &= \argmaxB_{\tilde{\boldsymbol{\theta}}}\langle \ln{p(\textbf{y},\textbf{x},\boldsymbol{\alpha},\bar{b};\tilde{\boldsymbol{\theta}}}) \rangle_{q(\textbf{x};\tilde{\boldsymbol{\theta}}^{OLD})q(\boldsymbol{\alpha};\tilde{\boldsymbol{\theta}}^{OLD})q(\bar{b};\tilde{\boldsymbol{\theta}}^{OLD})} \\
        &=\argmaxB_{\tilde{\boldsymbol{\theta}}}\langle \ln{p(\textbf{y}|\textbf{x};\tilde{\boldsymbol{\theta}}})p(\textbf{x}|\boldsymbol{\alpha};\tilde{\boldsymbol{\theta}}) \rangle_{q(\textbf{x};\tilde{\boldsymbol{\theta}}^{OLD})q(\boldsymbol{\alpha};\tilde{\boldsymbol{\theta}}^{OLD})}.
\end{split}
\end{equation*}
Let $Q(\tilde{\boldsymbol{\theta}}) = \langle \ln{p(\textbf{y}|\textbf{x};\tilde{\boldsymbol{\theta}}})p(\textbf{x}|\boldsymbol{\alpha};\tilde{\boldsymbol{\theta}}) \rangle_{q(\textbf{x};\tilde{\boldsymbol{\theta}}^{OLD})q(\boldsymbol{\alpha};\tilde{\boldsymbol{\theta}}^{OLD})}$, which results in
\begin{align}\label{eq:19}
      Q(B,\lambda) &= \langle \ln{p(\textbf{y}|\textbf{x};\lambda)} \rangle_{q(\textbf{x};\tilde{\boldsymbol{\theta}}^{OLD})q(\boldsymbol{\alpha};\tilde{\boldsymbol{\theta}}^{OLD})} 
      +\langle \ln{p(\textbf{x}|\boldsymbol{\alpha};\text{B})} \rangle_{q(\textbf{x};\tilde{\boldsymbol{\theta}}^{OLD})q(\boldsymbol{\alpha};\tilde{\boldsymbol{\theta}}^{OLD})}  \nonumber \\
      & =  \langle \ln{p(\textbf{y}|\textbf{x};\lambda)} \rangle_{q(\textbf{x};\tilde{\boldsymbol{\theta}}^{OLD})} 
      +\langle \ln{p(\textbf{x}|\boldsymbol{\alpha};\text{B})} \rangle_{q(\textbf{x};\tilde{\boldsymbol{\theta}}^{OLD})q(\boldsymbol{\alpha};\tilde{\boldsymbol{\theta}}^{OLD})}.  
\end{align}
$\text{B}$ and $\lambda$ are estimated as Eqs.~\eqref{eq:22} and~\eqref{eq:25}, respectively, by maximizing the Eq.~\eqref{eq:19}. 
\begin{equation}\label{eq:22}
    \text{B}= [\frac{1}{N}\sum_{i=1}^{N}\langle \alpha_{i} \rangle (\Sigma_{\text{x}_{i}} + \langle \text{x}_{i} \rangle \langle \text{x}_{i} \rangle^{\top})]^{-1}
\end{equation}
\begin{equation}\label{eq:25}
    \lambda=\frac{[\lVert \textbf{y}-\textbf{D} \mu_{\textbf{x}} \rVert_{2}^{2} + \hat{\lambda} [NL - \text{Tr}(\Sigma_{\textbf{x}}\mathbb{E}_{q(\boldsymbol{\alpha};\tilde{\boldsymbol{\theta}}^{OLD})}(\Sigma_{0}^{-1}))]]}{ML}.
\end{equation}
Detailed derivations of Eqs.~\eqref{eq:22} and~\eqref{eq:25} are described in  Appendices~\ref{app:app4} and~\ref{app:app5}, respectively.

Algorithm \ref{alg:alg1} shows the procedure of the proposed SA-TSBL method. Given the \textcolor{black}{multiple KPCs samples} ($\textbf{y}$) and fault pattern matrix ($\Phi$), the proposed method estimates the following variables and parameters in E and M steps, respectively.
\vspace{-0.3cm}\begin{itemize}
    \item E-step: Process errors ($\mu_{\textbf{x}}$), the variable related to the sparsity of process errors ($ \boldsymbol{\alpha}$), and the variable that is used to provide prior knowledge of process faults ($\bar{b}$).\vspace{-0.3cm}
    \item M-step: \textcolor{black}{Temporal correlations in the time series data of each KCC ($\text{B}$) and noise ($\lambda$).}
\end{itemize}\vspace{-0.3cm}
These steps iterate until the estimator of process errors ($\mu_{\textbf{x}}$) is rarely updated, namely, $\lVert \mu_{\textbf{x}}^{t-1}-\mu_{\textbf{x}}^{t} \rVert_{\infty}<\gamma$, where $\lVert \cdot \rVert_{\infty}$ indicates infinity norm and $\gamma$ is a user-defined threshold (e.g., $\gamma=10^{-6}$). \textcolor{black}{Then, the mean deviation of process errors, which is the output of Algorithm \ref{alg:alg1}, are calculated by the following procedure.}
\vspace{-0.3cm}\begin{itemize}
    \item Step 1: The matrix $\tilde{\mu}_{\textbf{x}}$ is defined in which the $i^{th}$ row represents Eq.~\eqref{eq:100} derived from $\mu_{\textbf{x}}$.\vspace{-0.3cm}
    \item Step 2: The average of each row of matrix $\tilde{\mu}_{\textbf{x}}$, that is, a vector  $\bar{\tilde{\mu}}_{\textbf{x}}$ is provided as the output of Algorithm \ref{alg:alg1}, indicating the mean deviation of process errors.
\end{itemize}\vspace{-0.2cm}
Therefore, nonzero values in a vector $\bar{\tilde{\mu}}_{\textbf{x}}$ are process faults of the mean shifts.


\vspace{-0.2cm}
\begin{algorithm}[!htb]\label{alg:alg1}
    \DontPrintSemicolon
    \textbf{Input}: \textcolor{black}{Multiple KPCs samples} ($\textbf{y}$), Fault pattern matrix ($\Phi$).\\
    \vspace{-0.2cm}
    \textbf{Set} $a=b_{i}\hspace{0.25em} (i \in P^{c})=10^{-4}$. $p$ and $q$ are set based on two conditions in Section~\ref{s:methods.3.1}.\\
    \vspace{-0.2cm}
    \textbf{Initialize}  $\text{B}=\textbf{I}_{L}, \lambda=1,  b_{i} \hspace{0.25em} (i\in P) =1,,t=1.$\\
    \vspace{-0.2cm}
    \textbf{While  $\lVert \mu_{\textbf{x}}^{t-1}-\mu_{\textbf{x}}^{t} \rVert_{\infty}< \gamma $ do}\\
    \vspace{-0.2cm}
    \hspace{2.50em} \textbf{E-step of VBEM}:\\
    \vspace{-0.2cm}
    \hspace{5.00em} Update $ \mu_{\textbf{x}}$ using Eq.~\eqref{eq:15}\\
    \vspace{-0.2cm}
    \hspace{5.00em} Update $ \boldsymbol{\alpha}$ using Eq.~\eqref{eq:16}\\
    \vspace{-0.2cm}
    \hspace{5.00em} Update $ \bar{b}$ using Eq.~\eqref{eq:17}\\
    \vspace{-0.2cm}
    \hspace{2.50em} \textbf{M-step of VBEM}:\\
    \vspace{-0.2cm}
    \hspace{5.00em} Update $\text{B}$ using Eq.~\eqref{eq:22}\\
    \vspace{-0.2cm}
    \hspace{5.00em} Update $\lambda$ using Eq.~\eqref{eq:25}\\
    \vspace{-0.20cm}
    \hspace{2.50em} $t=t+1$\\
    \vspace{-0.20cm}
    \textbf{End}\\
    \vspace{-0.20cm}
    \textbf{Output}: Mean deviations of process errors $\bar{\tilde{\mu}}_{\textbf{x}}.$
    \caption{Proposed SA-TSBL method}
\end{algorithm}\vspace{-0.2cm}
\section{Numerical Case Studies} \label{s:sec4}
This section provides three scenarios to compare the performance between the proposed method and benchmark methods. 
\vspace{-0.3cm}\begin{itemize}
    \item Section~\ref{s:4.1}  shows the performance evaluation by varying the temporal correlation. This study is to validate the effectiveness of strong temporal correlation in sparse estimation.\vspace{-0.3cm}
    \item Section~\ref{s:4.2} provides the numerical study to investigate the impact of \textcolor{black}{the number of KPCs samples (i.e., measurement samples)} on the performance of process faults identification.\vspace{-0.3cm}
    \item Section~\ref{s:4.3} illustrates the sparse estimation performance by varying the ratio between the number of measurements and process errors (i.e., the severity of the underdetermined systems). 
\end{itemize}\vspace{-0.3cm}
All the numerical case studies consist of 100 independent trials. $p$ and $q$ in Eq.~\eqref{eq:7} are determined as 1 and 0.1, respectively, in the case studies to satisfy the two conditions provided in Section~\ref{s:methods.3.1}. The code of the proposed SA-TSBL algorithm is implemented in Matlab 2017. The CPU of the computer used in this paper is an Intel\textsuperscript{\textregistered} Core\textsuperscript{\texttrademark} Processor i7-8750H.

\begin{figure}[ht]
    \centering
    \resizebox*{7cm}{!}{\includegraphics{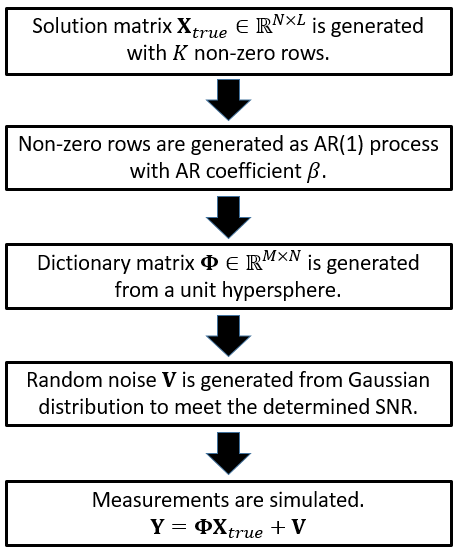}}
    \vspace{-0.5cm}\caption{Flowchart of the data generation process.}
    \label{fig:fig4}
\end{figure}

The benchmark methods selected in this study are as follows, which are widely used in sparse Bayesian learning.
\vspace{-0.3cm}\begin{itemize}
\item MSBL proposed in \cite{wipf2007empirical} is a basic SBL method for the MMV model that assumes \textcolor{black}{independence in the time series data of each KCC}.\vspace{-0.3cm}
\item T-MSBL proposed in \cite{zhang2011sparse} is a typical SBL method for the MMV model that considers \textcolor{black}{temporal correlation in the time series data of each KCC}. \vspace{-0.3cm}
\item SA-MSBL proposed in \cite{yu2019sa} is the SBL method that considers prior knowledge of support in the MMV model. It assumes \textcolor{black}{independence in the time series data of each KCC}.\vspace{-0.3cm}
\item SA-SBL proposed in \cite{fang2015support} and \cite{li2017adaptive} is the SBL method that considers prior knowledge of process faults in the single measurement vector (SMV) model. Like \cite{bastani2018fault}, the average of \textcolor{black}{multiple KPCs samples} is used to estimate the mean deviation of process errors.
\end{itemize}\vspace{-0.3cm}
\textbf{Data Generations}: The data generation process is summarized in Figure~\ref{fig:fig4}. The solution matrix $\textbf{X}_{true}\in \mathbb{R}^{N\times L}$ is randomly generated with $K$ nonzero rows. Indexes of the nonzero rows are randomly chosen in each trial. The nonzero rows in $\textbf{X}_{true}$ are generated as AR(1) process that initiates from the standard Gaussian distribution since AR(1) processes are sufficient to represent the temporal structure of the small number of measurement samples ($L$) \citep{zhang2011sparse}. The AR coefficient, defined as $\beta$, represents the temporal correlation. A dictionary matrix $\Phi \in \mathbb{R}^{M \times N}$ is constructed with columns drawn from the surface of a unit hyper-sphere uniformly \citep{donoho2006most}. Finally, the measurements matrix is built by $ \textbf{Y} =\Phi\textbf{X}_{true}+\textbf{V}$ in the final step of Figure~\ref{fig:fig4}, where $\textbf{V}$ is a Gaussian noise matrix with zero-mean \citep{zhang2011sparse}. The variance of the noise matrix is chosen to meet the determined value of the signal-to-noise ratio (SNR). SNR is defined by SNR(dB)$\vcentcolon= 20(\text{log}_{10} (\lVert \Phi \textbf{X}_{true} \rVert_{F} / \lVert \textbf{V} \rVert_{F}))$ \citep{zhang2011sparse}.
\vspace{-0.3cm}\\
\textbf{Performance Evaluation}: Since the objective of this paper is to diagnose the mean deviation of process errors, the target of the proposed method is to accurately estimate $\bar{\textbf{X}}_{true}$, which is the row-wise mean of $\textbf{X}_{true}$.

Two performance measures are used in this paper. One is the failure rate defined in \cite{li2017adaptive}. \textcolor{black}{It measures the accuracy of detecting process fault, which are the nonzero rows in $\bar{\textbf{X}}_{true}$. Assume that the number of process faults $K$ is given. Then, the row indexes of the $K$ largest $\ell_2$-norms from $\bar{\tilde{\mu}}_{\textbf{x}}$ in Algorithm \ref{alg:alg1} are identified. If the indexes are different from the indexes of nonzero rows in $\bar{\textbf{X}}_{true}$, it is considered a failed trial. The failure rate is the percentage of failed trials in the total trials.} 
The other performance measure is a normalized mean squared error (NMSE) that is defined by $\lVert \bar{\tilde{\mu}}_{\textbf{x}}-\bar{\textbf{X}}_{true} \rVert_{F}^{2} / \lVert \bar{\textbf{X}}_{true} \rVert_{F}^{2}$. Averages of failure rate and NMSE from 100 trials are used as performance measures.
\begin{remark}{ \textnormal{\textbf{(Performance Evaluations of SA-TSBL, SA-MSBL, and SA-SBL)}}}\label{r:r1}

The SA-MSBL, SA-SBL, and the proposed SA-TSBL have several cases based on the set of prior knowledge (i.e., $P$ in Eq.~\eqref{eq:7}) even in the same problem. Both correct and erroneous information of support exists in the set $P$ as prior knowledge. To differentiate between the correct and erroneous information of support, two subsets, namely, $P_{C}$ and $P_{E}$ are defined. $P_{C}$ consists of the indexes of nonzero rows obtained from prior knowledge among true nonzero rows in $\bar{\textbf{X}}_{true}$. In contrast, $P_{E}$ consists of the indexes of nonzero rows in prior knowledge but are actually zero in $\bar{\textbf{X}}_{true}$. The cardinalities of $P_{C}$ and $P_{E}$ in this paper are assumed to satisfy two conditions. The first condition is the cardinality of set $P_{C}$ and $P_{E}$ should be less than or equal to 75$\%$ and 50$\%$ of the number of nonzero rows in $\bar{\textbf{X}}_{true}$ (i.e., $K$), respectively. For example, if $K$ is 6, the cardinality of $P_{C}$ and $P_{E}$ are less than or equal to 4 and 3, respectively. The first condition illustrates the cardinality of prior knowledge is similar to $K$. \textcolor{black}{The condition enables the performance evaluation of the proposed method in comprehensive situations. Since the cardinality of $P_{C}$ is less than or equal to 4, the prior knowledge is partial (i.e., $P_{C}$ misses some rows that are actually nonzero in $\bar{\textbf{X}}_{true}$) in all cases. In addition, the cases with a cardinality of $P_{E}$ greater than zero show the situations when the erroneous prior knowledge (i.e., $P_{E}$ contains the rows that are actually zero in $\bar{\textbf{X}}_{true}$) exist}. The second condition is the cardinality of $P_{C}$ is greater than or equal to that of $P_{E}$. The second condition prevents erroneous prior knowledge from being dominant prior knowledge. Therefore, performances of these three methods are evaluated as the average of all cases with different cardinalities of $P_{C}$ and $P_{E}$, where each case consists of 100 trials.
\end{remark}
\subsection{\emph{Performance Evaluation in Various Temporal Correlations}} \label{s:4.1}
This case study shows the performance of all methods in various temporal correlations under the noiseless case. The size of the dictionary matrix $\Phi$ is 8$\times$40. The number of nonzero rows of $\bar{\textbf{X}}_{true}$ ($K$), and measurement samples ($L$) are 6 and 3, respectively. Temporal correlation ($\beta$) varies among 0.1, 0.3, 0.5, 0.7, 0.9, and 0.99. The proposed method shows the best performance in all temporal correlations, as shown in Table~\ref{tab: table1}.\begin{table}[!htb]
\centering
\caption{Performance comparison with various temporal correlations ($\beta$).}
\label{tab: table1}
\begin{tabular}{cccccc|ccccc}
\hline\hline
         & \multicolumn{5}{c}{Failure Rate} & \multicolumn{5}{c}{NMSE}         \\
\cline{2-11}          
 $\beta$   & 0.1  & 0.3  & 0.6  & 0.9  & 0.99 & 0.1  & 0.3  & 0.6  & 0.9  & 0.99 \\
\hline T-MSBL   & 0.35 & 0.36 & 0.29 & 0.22 & 0.20 & 0.62 & 0.54 & 0.42 & 0.35 & 0.26 \\
\hline MSBL     & 0.54 & 0.57 & 0.73 & 0.88 & 0.98 & 0.58 & 0.59 & 0.84 & 1.02 & 1.25 \\
\hline SA-MSBL  & 0.98 & 0.98 & 0.98 & 0.98 & 0.99 & 0.75 & 0.78 & 0.82 & 0.89 & 0.97 \\
\hline SA-SBL   & 0.99 & 0.99 & 0.99 & 0.99 & 0.99 & 0.84 & 0.86 & 0.90 & 0.91 & 0.99 \\ \hline
\begin{tabular}[c]{@{}c@{}}\textbf{SA-TSBL}\\ \vspace{-0.0cm}\textbf{(Proposed)}\end{tabular} & \textbf{0.27} & \textbf{0.27} & \textbf{0.23} & \textbf{0.18} & \textbf{0.16} & \textbf{0.45} & \textbf{0.40} & \textbf{0.34} & \textbf{0.28} & \textbf{0.22} \\ \hline \hline
\end{tabular}
\end{table}
\begin{figure}[!htb]
    \centering
    \resizebox{\textwidth}{!}{\includegraphics{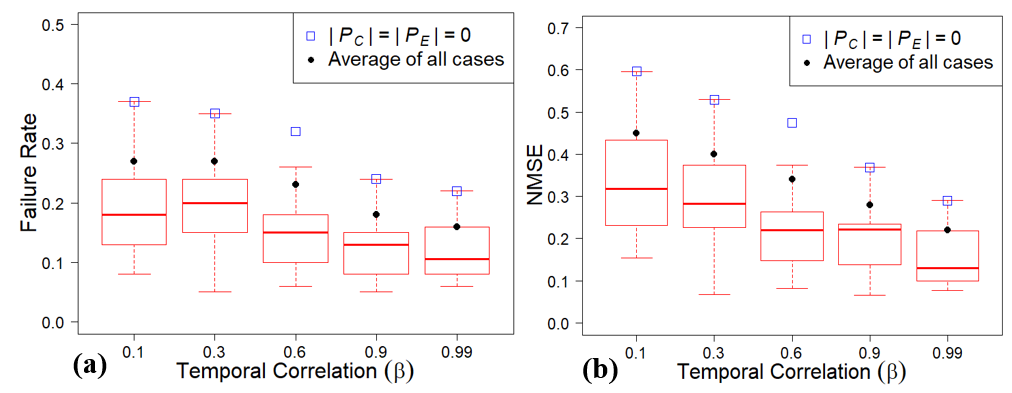}}\vspace{-0.5cm} 
    \caption{The boxplot for (a) failure rate and (b) NMSE in all cases of SA-TSBL in various temporal correlations.}
    \label{fig:fig5} \vspace{-0.3cm}
\end{figure}Specifically, the failure rate and NMSE of the proposed method and T-MSBL tend to decrease as $\beta$ increases since these two methods capture the temporal correlation. In contrast, both measures of MSBL and SA-MSBL show the opposite trend as the two methods assume independence among samples. 

The performance evaluations of the proposed method with different cardinalities of $P_{C}$ and $P_{E}$ are represented as boxplots in Figure~\ref{fig:fig5}. Since the number of nonzero rows is 6, each correlation has 14 cases designed under the two conditions of Remark \ref{r:r1}. In each boxplot, the horizontal line represents the minimum, the first quartile, median, the third quartile, and maximum value sequentially from the bottom, respectively. The minimum value denotes the smallest value within the 1.5 $\times$ interquartile range (IQR) below the first quartile. Similarly, the maximum value is defined as the largest value that is within the 1.5 $\times$ IQR above the third quartile \citep{hubert2008adjusted}. A blue square in Figure~\ref{fig:fig5} shows the performance of the proposed method without any prior knowledge (i.e., $|P_{C}|=|P_{E}|=0$), and the black dot represents the average of all cases used as a performance evaluation measure of the proposed method. The trend of black dots shows the performance improvements of the proposed method as  temporal correlation ($\beta$) increases. In addition, boxplots in Figure~\ref{fig:fig5} represent all cases with partial and some erroneous prior knowledge achieve better performance in both measures than those without prior knowledge. Even in the case of $\beta=0.6$, a square is considered an outlier since it is located higher than 
 1.5 $\times$ IQR from the third quartile of the boxplot. The results show the effectiveness of prior knowledge in the proposed method, which successfully distinguishes the correct and incorrect prior knowledge in various temporal correlations.

\subsection{\emph{
Performance Evaluation in the Various Numbers of Measurement Samples 
}} \label{s:4.2}
The objective of this case study is to compare the performance of all methods using the various numbers of measurement samples ($L$). The size of the dictionary matrix $\Phi$ is 7$\times$55, and the number of nonzero rows in $\bar{\textbf{X}}_{true}$ ($K$) is 4. SNR and temporal correlation ($\beta$) are set as 35dB, and 0.95, respectively.
 The number of measurement samples varies from 2 to 4. Table~\ref{tab:table2} shows that the proposed method achieves the best performance in all cases. \vspace{-0.0cm}\begin{table}[!ht]
\centering
\caption{Performance comparison with various number of measurement samples ($L$).}
\label{tab:table2}
\begin{tabular}{cccc|ccc}
\hline\hline
         & \multicolumn{3}{c}{Failure   Rate} & \multicolumn{3}{c}{NMSE} \\
         \cline{2-7}
$L$        & 2          & 3         & 4         & 2      & 3      & 4      \\
\hline T-MSBL   & 0.59       & 0.17      & 0.13      & 0.75   & 0.19   & 0.13   \\
\hline MSBL     & 0.83       & 0.77      & 0.52      & 1.09   & 1.04   & 0.64   \\
\hline SA-MSBL  & 0.86       & 0.88      & 0.89      & 1.04   & 1.09   & 0.92   \\
\hline SA-SBL   & 0.88       & 0.89      & 0.89      & 1.06   & 1.05   & 0.92   \\\hline
\begin{tabular}[c]{@{}c@{}}\textbf{SA-TSBL}\\ \vspace{-0.0cm}\textbf{(Proposed)}\end{tabular}  & \textbf{0.45}       & \textbf{0.12}      & \textbf{0.10}      & \textbf{0.52}   & \textbf{0.12}   & \textbf{0.10}  \\ \hline \hline
\end{tabular}
\end{table}\vspace{-0.0cm} In addition, the results in Table~\ref{tab:table2} illustrate that most of the MMV models achieve performance improvements as the number of measurement samples increases because of common support assumption \citep{cotter2005sparse}.

\begin{figure}[!ht]
    \centering
    \resizebox{\textwidth}{!}{\includegraphics{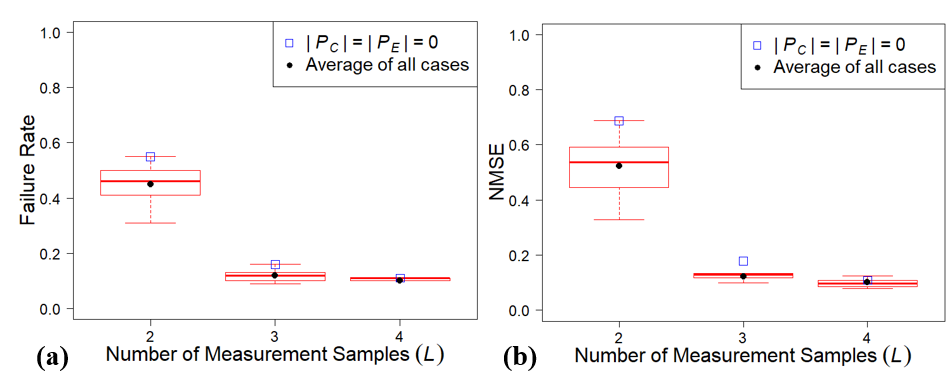}}\vspace{-0.5cm}
    \caption{The boxplot for (a) failure rate and (b) NMSE in all cases of SA-TSBL in various numbers of measurement samples.}
    \label{fig:fig6} \vspace{-0.3cm}
\end{figure}
Figure~\ref{fig:fig6} shows the performance evaluation of the proposed method with various sizes of $P_{C}$ and $P_{E}$. Especially, the performance of T-MSBL and the proposed method without prior knowledge (blue square in Figure~\ref{fig:fig6}) is similar. However, the proposed achieves better performance than T-MSBL by utilizing partial and some erroneous prior knowledge. Specifically, the performance of the proposed method improves from 4$\%$ to 31$\%$ of those without any prior knowledge (a relative improvement from square to dot in Figure~\ref{fig:fig6}) in the various numbers of \textcolor{black}{measurement samples.}
\subsection{\emph{Performance Evaluation in Various Ratios between the Number of Measurements and Process Errors}} \label{s:4.3}
\begin{figure}[ht]
    \centering
    \resizebox{\textwidth}{!}{\includegraphics{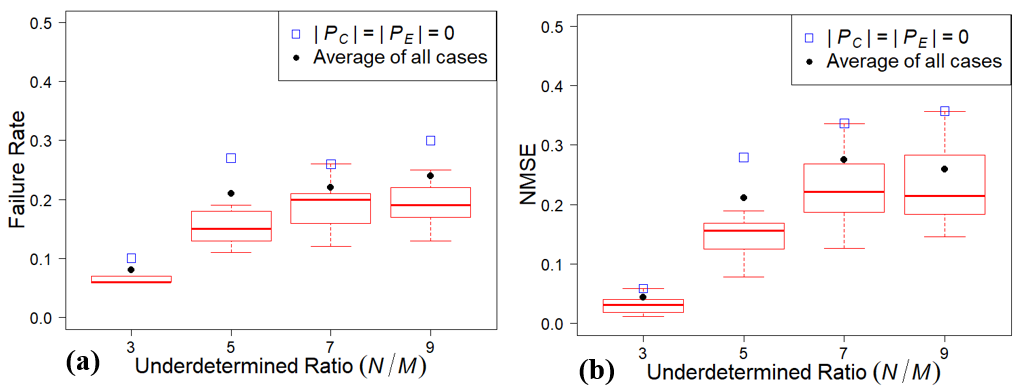}}\vspace{-0.5cm}
    \caption{ The boxplot for (a) failure rate and (b) NMSE in all cases of SA-TSBL in various underdetermined ratio.}
    \label{fig:fig7}\vspace{-0.3cm}
\end{figure}

Results in Table~\ref{tab:table3} illustrate the performance of all methods by varying the ratio between the number of measurements and process errors (i.e., underdetermined ratio). In this study, the number of measurements ($M$) is fixed at 10, and the underdetermined ratio ($N/M$) is selected from 3, 5, 7, and 9 with 25dB for SNR, respectively. The number of nonzero rows in $\bar{\textbf{X}}_{true}$ ($K$) and measurement samples ($L$) are set as 4, 3, respectively. The temporal correlation ($\beta$) is 0.99. As the underdetermined ratio increases, it becomes more challenging to identify the sparse process faults. However, capturing temporal correlation and utilizing the partial and some erroneous prior knowledge enable the proposed method to achieve the best performance even in a high underdetermined ratio. Figure~\ref{fig:fig7} shows prior knowledge of support is still valuable in various underdetermined ratios. This study shows that the proposed algorithm can be applied to applications such as neuroimaging in that highly underdetermined systems exist. 
\begin{table}[!htbp]\vspace{-0.5cm}
\centering
\caption{Performance comparison with various underdetermined ratio ($N/M$).}
\label{tab:table3}
 \begin{tabular}{ccccc|cccc}
\hline\hline
                  & \multicolumn{4}{c}{Failure Rate}                              & \multicolumn{4}{c}{NMSE}                                      \\ \cline{2-9} 
$N/M$  & 3&5 &7 & 9   &3  &5  &7  &9  \\
\hline T-MSBL  & 0.09  & 0.28  & 0.26  & 0.31  & 0.05  & 0.31  & 0.33  & 0.42 \\
\hline MSBL & 0.29 & 0.56 & 0.64  & 0.69  & 0.30  & 0.61 & 0.77  & 0.85\\
\hline SA-MSBL & 0.42 & 0.66 & 0.73 & 0.82 & 0.35 & 0.59 & 0.65 & 0.73 \\
\hline SA-SBL & 0.67 & 0.85 & 0.88 & 0.90 & 0.48 & 0.67 & 0.71 & 0.75  \\\hline
\begin{tabular}[c]{@{}c@{}}\textbf{SA-TSBL}\\ \vspace{-0.0cm}\textbf{(Proposed)}\end{tabular} & \textbf{0.08} & \textbf{0.21} & \textbf{0.22} & \textbf{0.24} & \textbf{0.04} & \textbf{0.21} & \textbf{0.28} & \textbf{0.29} \\ \hline \hline
\end{tabular}
\end{table}
\vspace{-0.3cm}
\section{Real-World Simulation Case Studies}\label{s:sec5}
\begin{figure}[!htbp]
    \centering
    \resizebox{0.7\textwidth}{!}{\includegraphics{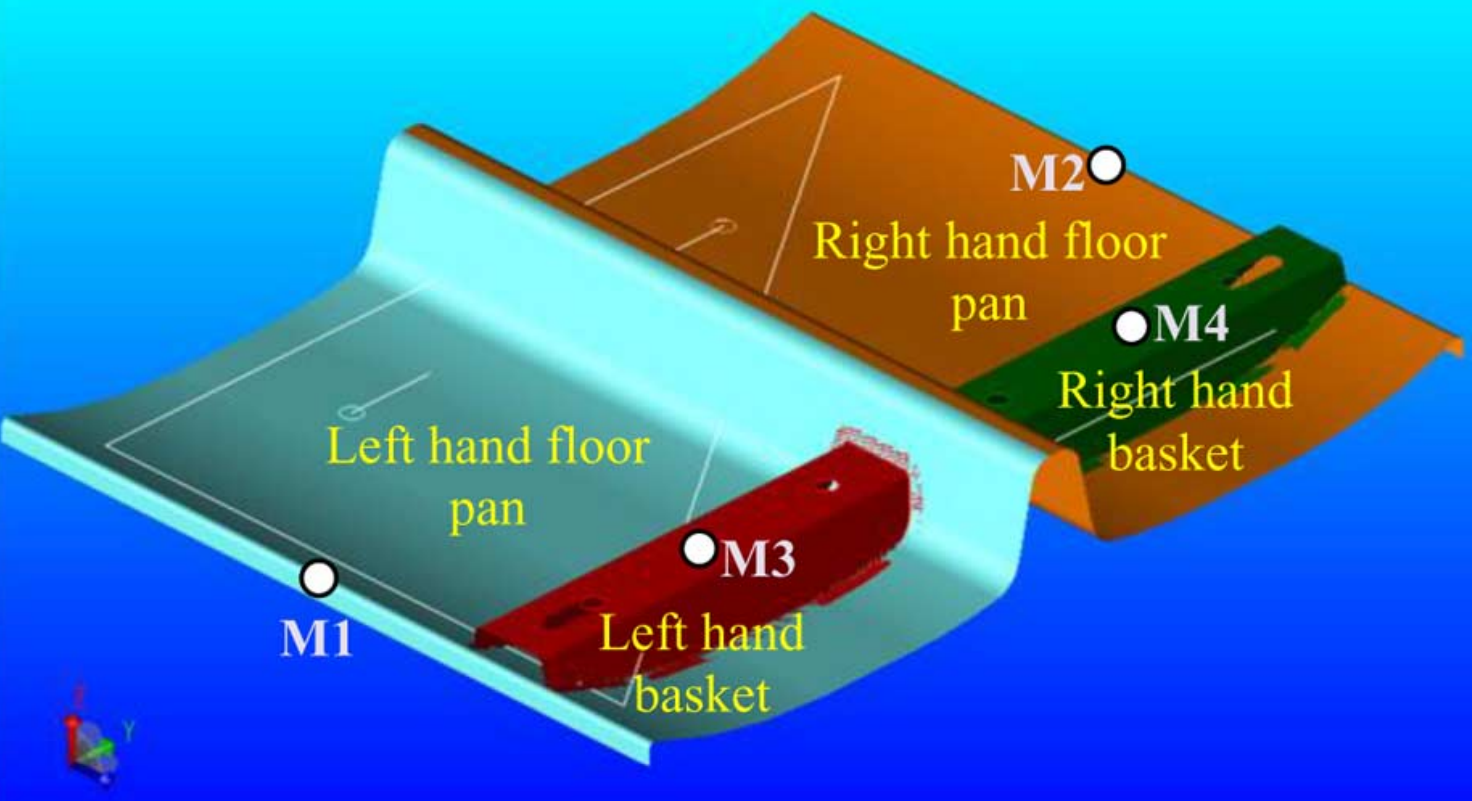}}\vspace{-0.5cm}
    \caption{Floor-pan assembly model \citep{bastani2012fault}.}
    \label{fig:fig8}\vspace{-0.3cm}
\end{figure}

\begin{figure}[!htbp]
    \centering
    \resizebox{0.7\textwidth}{!}{\includegraphics{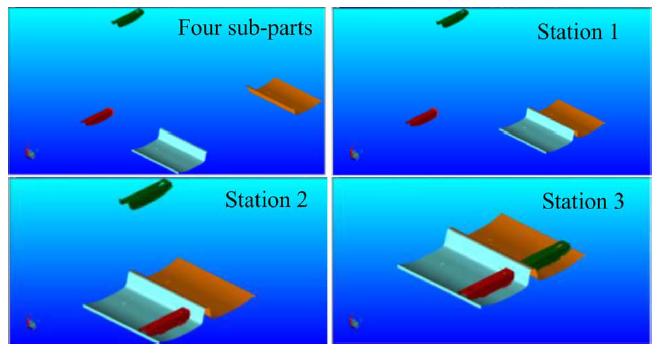}}\vspace{-0.5cm}
    \caption{Floor-pan assembly procedure from three assembly stations \citep{bastani2012fault}.}
    \label{fig:fig999}
\end{figure}

An assembly model from a real auto body assembly process is utilized as a real-world case study. As shown in Figure~\ref{fig:fig8}, the assembled product is a floor pan of a car. \textcolor{black}{It consists of four parts namely, the left floor pan, left bracket, right floor pan, and right bracket.} 
\textcolor{black}{They are assembled in three stations as illustrated in Figure~\ref{fig:fig999}. During the assembly process, the parts are held by fixtures and part-mating features, which are the KCCs in this process \citep{bastani2016compressive}.} KPCs are measured from four points, namely, M1, M2, M3, and M4, respectively, as shown in Figure~\ref{fig:fig8}. \textcolor{black}{Each part has its own measurement place such as M1 on part 1, M2 on part 2, M3 on part 3, and M4 on part 4. These measurements can be measured in each station when the corresponding part has been assembled in the previous stations. For example, M4 on part 4 cannot be measured in station 1 because part 4 has not yet been assembled at station 1 \citep{bastani2016compressive}.} In addition, KPCs are measured in three directions (X, Y, and Z) at each point. In this assembly process, there exists a total of 33 process errors, which are fixture locator dimensional errors \citep{bastani2018fault}. \textcolor{black}{The dimension of the fault pattern matrix is 12$\times$33, and is established based on the literature \citep{bastani2016compressive,huang2007stream,kong2009variation}. Since the variations of KCCs can be propagated to deviations of other KCCs in the subsequent stations, the transfer and accumulation of KCCs deviations between multiple stations are considered in the formulation of the fault pattern matrix $\Phi$ \citep{ding2000modeling,bastani2016compressive}.} The matrix is provided in Appendix \ref{app:app6}. Since the number of measurements (12) is less than the number of process errors (33), it causes an underdetermined system in the fault quality linear model. It requires sparse estimation to identify process faults.

\vspace{-0.0cm}

\textcolor{black}{To generate the multiple KPCs samples from the auto body assembly process, the time series data of each process fault is generated from AR (1) process initiated from the standard Gaussian distribution as in Section~\ref{s:sec4}. The generated temporal correlated process faults ($\textbf{X}_{true}$) and fault pattern matrix provide the multiple KPCs samples ($\textbf{Y}$), as in Figure~\ref{fig:fig4}. In addition, performance evaluation measures,  benchmark methods, and the values of $p$ and $q$ in Eq.~\eqref{eq:7} used in Section~\ref{s:sec4} are still utilized in Section~\ref{s:sec5}. Prior knowledge of process faults is also provided in the same way as Section~\ref{s:sec4}.} Sections \ref{s:methods.5.1} and \ref{s:methods.5.2} show the performance evaluation \textcolor{black}{by varying the number of KPCs samples, and temporal correlation $\beta$, respectively.}

\subsection{\emph{
Performance Evaluation in the Various Number of KPCs Samples 
}} \label{s:methods.5.1}
This case study aims to demonstrate the effectiveness of sparse estimation of process faults by varying the number of \textcolor{black}{KPCs samples ($L$) when a strong correlation exists ($\beta$=0.99). The number of KPCs samples} varies from 2 to 4, and SNR is 50dB. Three process faults ($K$) are determined among 33 process errors randomly. The proposed method shows the best performance in all various numbers of \textcolor{black}{KPCs samples}, as shown in Table~\ref{tab:table4}.\begin{table}[!htbp]\vspace{-0.5cm}
\centering
\caption{Performance comparison by varying the number of KPCs samples.}
\label{tab:table4}
\begin{tabular}{cccc|ccc}
\hline\hline
         & \multicolumn{3}{c}{Failure Rate} & \multicolumn{3}{c}{NMSE}  \\
         \cline{2-7} 
KPCs samples              & 2      & 3      & 4        & 2    & 3    & 4    \\
\hline T-MSBL     & 0.59   & 0.63   & 0.51  & 0.34 & 0.32 & 0.25 \\
\hline MSBL        & 0.59   & 0.62   & 0.50   & 0.36 & 0.33 & 0.24 \\
\hline SA-MSBL     & 0.57   & 0.62   & 0.52   & 0.40 & 0.40 & 0.34 \\
\hline SA-SBL     & 0.60   & 0.65   & 0.58   & 0.43 & 0.42 & 0.40 \\\hline
\begin{tabular}[c]{@{}c@{}}\textbf{SA-TSBL}\\ \textbf{(Proposed)}\end{tabular}     & \textbf{0.48}   & \textbf{0.54}   & \textbf{0.38}  & \textbf{0.30} & \textbf{0.30} & \textbf{0.21} \\ \hline \hline
\end{tabular}
\end{table} All methods except SA-SBL generally tend to improve performance in both measures as the number of KPCs samples increases. However, the performances of all methods do not improve significantly as \textcolor{black}{the number of KPCs samples} increases, compared to Section~\ref{s:4.2}. The reason could be the fault pattern matrix in this multistation assembly process, which is very structured compared to the random design matrix $\Phi$ in Section~\ref{s:sec4}. \textcolor{black}{The random matrix achieves an accurate sparse estimation based on some theoretical properties of the random matrix, such as low mutual coherence that measures the highest correlation between columns of $\Phi$ \citep{candes2008introduction}.} \textcolor{black}{In contrast, if the mutual coherence is high, it causes inaccurate sparse estimation \citep{elad2007optimized}. As shown in Appendix~\ref{app:app6}, the columns of fault pattern matrix $\Phi$ in the assembly process are highly correlated, and mutual coherence is 1, causing the challenging sparse estimation task than Section~\ref{s:4.2}. Therefore, the results in this section are not significantly improved compared to Section~\ref{s:4.2}, which uses a random matrix with low mutual coherence, even if the number of KPCs samples is increased.}

\begin{figure}[!htbp]
    \centering
    \resizebox{\textwidth}{!}{\includegraphics{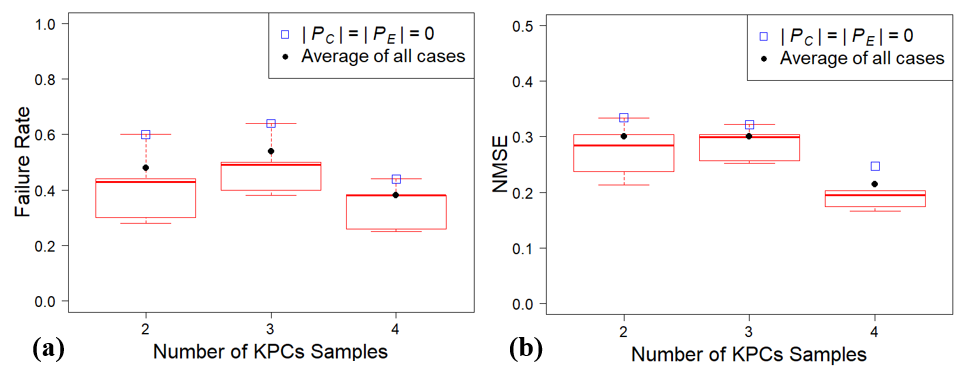}}\vspace{-0.5cm}
    \caption{ The boxplot for (a) failure rate and (b) NMSE in all cases of SA-TSBL in the various number of KPCs samples.}
    \label{fig:fig9}\vspace{-0.3cm}
\end{figure}

Figure~\ref{fig:fig9} shows that utilizing prior knowledge of process faults is still effective in the proposed method for failure rate and NMSE, even in the structured design matrix $\Phi$. The property lets the proposed method obtain a more accurate sparse estimation than T-MSBL, MSBL that cannot incorporate the prior knowledge of process faults. 

\subsection{\emph{Performance Evaluation in Various Temporal Correlations}} \label{s:methods.5.2}
\begin{figure}[!htb]
    \centering
    \resizebox{\textwidth}{!}{\includegraphics{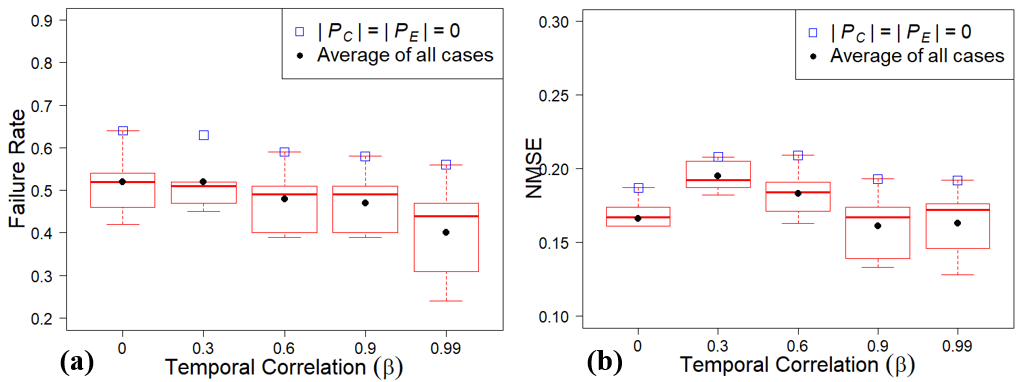}}\vspace{-0.5cm}
    \caption{The boxplot for (a) failure rate and (b) NMSE in all cases of SA-TSBL in various temporal correlations.}
    \label{fig:fig10}
\end{figure}

This case study presents the performance of all methods in various \textcolor{black}{temporal correlations $\beta$, when there exist five KPCs samples ($L$).} Temporal correlation varies among 0.0, 0.3, 0.6, 0.9, and 0.99. The study has three process faults ($K$) with noise level 80dB. Table~\ref{tab:table5} shows the proposed method achieves the best performance, and the performance of other methods exhibit similar trends to the case in Section. \ref{s:4.1}. The performance of the proposed method and T-MSBL improves in general as $\beta$ increases, and MSBL and SA-MSBL show the opposite trends since they assume the independence \textcolor{black}{in the time series data of each process error}. Figure~\ref{fig:fig10} illustrates the prior knowledge of process faults is still effective in various temporal correlations even with the structured matrix $\Phi$. 
\vspace{-0.3cm}\begin{table}[!htb]
\centering
\caption{ Performance comparison by varying temporal correlations ($\beta$).}\label{tab:table5}
\begin{tabular}{cccccc|ccccc} 
\hline\hline
                  & \multicolumn{5}{c}{Failure Rate}                                              & \multicolumn{5}{c}{NMSE}  \\\cline{2-11} 
    $\beta$              & 0.0           & 0.3           & 0.6           & 0.9           & 0.99          & 0.0           & 0.3           & 0.6           & 0.9           & 0.99          \\
\hline T-MSBL            & 0.62          & 0.59          & 0.56          & 0.52          & 0.54          & 0.19          & 0.21          & 0.21          & 0.19          & 0.20          \\
\hline MSBL              & 0.59          & 0.55          & 0.58          & 0.61          & 0.63          & 0.18          & 0.20          & 0.21          & 0.24          & 0.30          \\
\hline SA-MSBL           & 0.57          & 0.58          & 0.55          & 0.57          & 0.55          & 0.29          & 0.30          & 0.29          & 0.31          & 0.35          \\
\hline SA-SBL            & 0.75          & 0.73          & 0.72          & 0.66          & 0.60          & 0.46          & 0.45          & 0.42          & 0.38          & 0.39          \\
\hline
\begin{tabular}[c]{@{}c@{}}\textbf{SA-TSBL}\\ \vspace{-0.0cm}\textbf{(Proposed)}\end{tabular} & \textbf{0.52} & \textbf{0.52} & \textbf{0.48} & \textbf{0.47} & \textbf{0.40} & \textbf{0.17} & \textbf{0.20} & \textbf{0.18} & \textbf{0.16} & \textbf{0.16} \\ \hline \hline
\end{tabular}
\end{table}

\vspace{-0.3cm}\vspace{-0.0cm}
\section{Conclusions}\label{s:sec6}
This paper proposes a novel sparse hierarchical Bayesian method, SA-TSBL, to effectively identify the sparse process faults in multistation assembly systems. The method identifies process faults by considering the temporal correlation of each KCC and utilizing partial with some erroneous prior knowledge of process faults. Since posterior distributions of process errors in the proposed method are computationally intractable, this paper derives approximate posterior distributions of process errors via Variational Bayes inference. The effectiveness of the proposed method is validated by both numerical cases and real-world simulation application that uses an actual auto body assembly system. Based on these studies, it is evident that the direct use of multiple KPCs samples in the proposed SA-TSBL is more effective to process mean shift identification than in a previous study using the average of multiple KPCs samples in the single measurement vector model (Eq.~\eqref{eq:1}). This is because the proposed method can fully utilize multiple KPCs samples without any information loss. In addition, the results in case studies represent the proposed method achieves high performance in process faults estimation when \textcolor{black}{the time series data of each KCC} have a strong temporal correlation. Furthermore, utilizing the prior knowledge of process faults improves the performance of the proposed method in the case studies, even if the knowledge has some erroneous information. This is possible through the Bayesian framework in the proposed method that distinguishes between the correct and incorrect prior knowledge. 

\textcolor{black}{In this work, all fixture locators are assumed to be independent, which is a common assumption in the literature \citep{lee2020variation, bastani2016compressive, li2016bayesian}. However, there exists a correlation between process errors in some manufacturing systems. Specifically, the fixture locators in the multistation assembly process have some spatial correlation with each other if the locators are in the composite tolerance mode \citep{kong2006mode}. Therefore, expanding the proposed SA-TSBL considering the spatial correlation among the process errors is a promising direction for future work. In addition, instead of utilizing the Gamma distribution in Eq.~\eqref{eq:6} enforcing the sparsity of process error in the proposed method, other distributions such as  Laplacian distribution \citep{park2008bayesian}, and Generalized Pareto distribution \citep{zhang2017variational} can also be considered as prior distribution since the probability of these distributions is concentrated at zero. Based on each of these distributions, deriving new hierarchical models considering both the temporal correlation of each process error and prior knowledge of process faults is also a valuable direction of future work. Finally, extending the proposed method to consider the different temporal correlations between the process errors is an additional direction for future work.}


\bibliographystyle{chicago}
\spacingset{1}
\bibliography{IISE-Trans}


\spacingset{1.5} 
\appendix
\newpage
 \setcounter{page}{1}
\section{Appendix}
In this section, the approximate posterior distributions of hidden variables which are $\textbf{x}, \boldsymbol{\alpha}$, and $\bar{b}$ are derived via Variational Bayes inference. In addition, the derivations of hyperparameters by maximizing the complete likelihood in Eq. \eqref{eq:19} are also described.

\subsection{\emph{Inference for Eq.~\eqref{eq:12} }}\label{app:app1}
Based on Eq.~\eqref{eq:9}, $\ln{q(\textbf{x}})=\langle \ln{p(\textbf{y}|\textbf{x};\lambda})p(\textbf{x}|\boldsymbol{\alpha};\text{B}) \rangle_{q(\boldsymbol{\alpha})}+const.$ Therefore,
\begin{align}
    \ln{q(\textbf{x}}) &\propto \langle \ln{p(\textbf{y}|\textbf{x};\lambda})p(\textbf{x}|\boldsymbol{\alpha};\text{B}) \rangle_{q(\boldsymbol{\alpha})} \nonumber\\
    & = \langle \ln{(N(\textbf{y}|\textbf{Dx},\lambda \textbf{I}_{ML})}\prod_{i=1}^{N}N(\text{x}_{i}|\textbf{0},(\alpha_{i}\text{B})^{-1}) \rangle_{q(\boldsymbol{\alpha})}  \nonumber\\
    & \propto \langle -\frac{1}{2\lambda}(\textbf{y}-\textbf{Dx})(\textbf{y}-\textbf{Dx})^{\top}-\frac{1}{2}\textbf{x}[\text{AB}]\textbf{x} \rangle_{q(\boldsymbol{\alpha})} \nonumber
\end{align}
where $\text{AB}= \text{diag}[\alpha_{1}\text{B},...,\alpha_{N}\text{B}]$. Therefore, Eq.~\eqref{eq:12} is derived, and $q(\textbf{x})$ follows Gaussian distribution as follows:
\begin{equation*}
    q(\textbf{x})\sim N(\textbf{x}|\mu_{\textbf{x}},\Sigma_{\textbf{x}}),
\end{equation*}
where $\mu_{\textbf{x}}=\frac{1}{\lambda}\Sigma_{\textbf{x}}\textbf{D}^{\top}\textbf{y}$, $\Sigma_{\textbf{x}}=(\frac{1}{\lambda}\textbf{D}^{\top}\textbf{D}+\langle \text{AB} \rangle )^{-1}$.

\subsection{\emph{Inference for Eq.~\eqref{eq:13} }}\label{app:app2}
Based on Eq.~\eqref{eq:10}, $\ln{q(\boldsymbol{\alpha})}=\langle \ln{p(\textbf{x}|\boldsymbol{\alpha};\text{B})p(\boldsymbol{\alpha}|\bar{b})} \rangle_{q(\textbf{x})q(\bar{b})} +const$. Therefore,
\begin{align}
\ln{q(\boldsymbol{\alpha})} &\propto \langle \ln{p(\textbf{x}|\boldsymbol{\alpha};\text{B})p(\boldsymbol{\alpha}|\bar{b})}  \rangle_{q(\textbf{x})q(\bar{b})} \nonumber \\
&=\langle \ln{(\prod_{i=1}^{N}N(\text{x}_{i}|0,(\alpha_{i}\text{B})^{-1})  \prod_{i=1}^{N}\text{Gamma}(\alpha_{i}|a,b_{i})}\rangle_{q(\textbf{x})q(\bar{b})}  \nonumber\\
&\propto \langle \Sigma_{i=1}^{N}\ln{\alpha_{i}}^{\frac{L}{2}} - (\frac{\alpha_{i}}{2}\text{x}_{i}^{\top}\text{B}\text{x}_{i})+\ln{\alpha}_{i}^{a-1}-b_{i}\alpha_{i} \rangle_{q(\textbf{x})q(\bar{b})} \nonumber\\
& \propto \ln{\alpha}_{i}^{\frac{L}{2}+a-1}-\frac{\alpha_{i}}{2} \langle \text{x}_{i}^{\top}\text{B} \text{x}_{i} \rangle - \langle b_{i} \rangle \alpha_{i}, \nonumber 
\end{align}
where
    \begin{align}
        \langle \text{x}_{i}^{\top}\text{B}\text{x}_{i} \rangle & = \langle \text{Tr}(\text{x}_{i}^{\top}\text{B}\text{x}_{i}) \rangle  \nonumber \\
        & = \langle \text{Tr}(\text{x}_{i}^{\top}\text{x}_{i}\text{B}) \rangle \nonumber \\
        & = \text{Tr}(\langle \text{x}_{i}^{\top}\text{x}_{i} \rangle \text{B}) \nonumber \\
        &= \text{Tr}[( \Sigma_{\text{x}_{i}} + \langle \text{x}_{i} \rangle \langle \text{x}_{i}\rangle^{\top})\text{B}]. \nonumber
    \end{align}
Therefore, Eq.~\eqref{eq:13} is derived, and $q(\alpha_{i}), i\in P$ follows Gamma distribution as follows: 
\begin{equation*}
    q(\alpha_{i})=\text{Gamma}(\alpha_{i}|\frac{L}{2}+a,\frac{\text{Tr}[( \Sigma_{\text{x}_{i}} + \langle \text{x}_{i} \rangle \langle \text{x}_{i}\rangle^{\top})\text{B}]}{2} + \langle b_{i} \rangle).
\end{equation*}
For $q(\alpha_{i}), i\in P^{c}$ follows Gamma distribution as follows: 
\begin{equation*}
    q(\alpha_{i})=\text{Gamma}(\alpha_{i}|\frac{L}{2}+a,\frac{\text{Tr}[( \Sigma_{\text{x}_{i}} + \langle \text{x}_{i} \rangle \langle \text{x}_{i}\rangle^{\top})\text{B}]}{2} +  b_{i} ).
\end{equation*}
\subsection{\emph{Inference for Eq.~\eqref{eq:14} }}\label{app:app3}
Based on Eq.~\eqref{eq:11}, $\ln{q(\bar{b})}=\langle \ln{p(\boldsymbol{\alpha}|\bar{b})}p(\bar{b}) \rangle_{q(\textbf{x})q(\boldsymbol{\alpha})} +const.$ Therefore,
\begin{align}
\ln{q(\bar{b})}& \propto \langle \ln{p(\boldsymbol{\alpha}|\bar{b})}p(\bar{b}) \rangle_{q(\textbf{x})q(\boldsymbol{\alpha})} \nonumber\\
&= \langle \ln{\prod_{i\in P}\text{Gamma}(\alpha_{i}|a,b_{i})} \prod_{i\in P}\text{Gamma}(b_{i}|p,q) \rangle_{q(\textbf{x})q(\boldsymbol{\alpha})}   \nonumber \\
& \propto \Sigma_{i\in P }(-b_{i}\langle \alpha_{i} \rangle +a\ln{b_{i} + (p-1)\ln{b_{i}} -q b_{i}})  \nonumber \\
&=\Sigma_{i\in P}((p+a-1)\ln{b_{i}}-(q+\langle \alpha_{i} \rangle )b_{i}). \nonumber
\end{align}
Therefore Eq.~\eqref{eq:14} is derived, and $q(b_{i}), \forall i \in P$ follows Gamma distribution as follows: 
\begin{equation*}
    q(b_{i})=\text{Gamma}(b_{i}|p+a,\langle \alpha_{i} \rangle+q), \forall i \in P.
\end{equation*}

\subsection{\emph{Inference for Eq.~\eqref{eq:22} }}\label{app:app4}
To estimate \text{B}, let the second term in Eq.~\eqref{eq:19} as follows:
\begin{equation*}
    Q(\text{B})=\langle \ln{p(\textbf{x}|\boldsymbol{\alpha};\text{B}}) \rangle_{{q(\textbf{x};\tilde{\boldsymbol{\theta}}^{OLD})q(\boldsymbol{\alpha};\tilde{\boldsymbol{\theta}}^{OLD})}}.
\end{equation*}
Let $\Gamma$=diag($\alpha_{1}^{-1},...,\alpha_{N}^{-1}$). It can be shown that
\begin{align}\label{eq:20}
     \ln{p(\textbf{x}|\boldsymbol{\alpha};\text{B})} &= -\frac{1}{2}\ln{(|\Gamma^{-1}|^{L} |\text{B}^{-1} |^{N}   }) -\frac{1}{2}\textbf{x}^{\top}(\Gamma \otimes \text{B}) \textbf{x} \nonumber \\
     &=-\frac{1}{2}\ln{|\Gamma^{-1}|^{L}}-\frac{1}{2}\ln{|\text{B}^{-1}|^{N}}-\frac{1}{2}\textbf{x}^{\top}(\Gamma \otimes \text{B}) \textbf{x}.
\end{align}
$Q(\text{B})$  can be calculated by taking the expectation to Eq.~\eqref{eq:20}. Then, taking derivative $Q(\text{B})$  with respect to $\text{B}$ leads to
\begin{equation}\label{eq:21}
    \frac{\partial Q(\text{B})}{\partial \text{B}}=\frac{N}{2}\text{B}^{-1}-\frac{1}{2}\sum_{i=1}^{N}\langle \alpha_{i} \rangle (\Sigma_{\text{x}_{i}} + \langle \text{x}_{i} \rangle \langle \text{x}_{i} \rangle^{\top}).
\end{equation}
By letting Eq.~\eqref{eq:21} equals to zero, \text{B} is estimated as follows: 
\begin{equation*}
    \text{B}= [\frac{1}{N}\sum_{i=1}^{N}\langle \alpha_{i} \rangle (\Sigma_{\text{x}_{i}} + \langle \text{x}_{i} \rangle \langle \text{x}_{i} \rangle^{\top})]^{-1}
\end{equation*}
\subsection{\emph{Inference for Eq.~\eqref{eq:25} }}\label{app:app5}
To estimate $\lambda$, let the first term in Eq.~\eqref{eq:19} as follows:
\begin{equation*}
    Q(\lambda)=\langle \ln{p(\textbf{y}|\textbf{x};\lambda)} \rangle_{q(\textbf{x};\tilde{\boldsymbol{\theta}}^{OLD})q(\boldsymbol{\alpha};\tilde{\boldsymbol{\theta}}^{OLD})}. 
\end{equation*}
It can be shown that
\begin{align}
Q(\lambda)& \propto -\frac{ML}{2}\ln{\lambda}-\frac{1}{2\lambda}\mathbb{E}_{q(\textbf{x};\tilde{\boldsymbol{\theta}}^{OLD})q(\boldsymbol{\alpha};\tilde{\boldsymbol{\theta}}^{OLD})}\lVert \textbf{y}-\textbf{D} \textbf{x} \rVert_{2}^{2} \nonumber \\
& = -\frac{ML}{2}\ln{\lambda} -\frac{1}{2\lambda}\mathbb{E}_{q(\textbf{x};\tilde{\boldsymbol{\theta}}^{OLD})q(\boldsymbol{\alpha};\tilde{\boldsymbol{\theta}}^{OLD})}[\lVert \textbf{y}-\textbf{D}\mu_{\textbf{x}} \rVert_{2}^{2} + \text{Tr}(\Sigma_{\textbf{x}}\textbf{D}^{\top}\textbf{D})] \nonumber \\
& = -\frac{ML}{2}\ln{\lambda} -\frac{1}{2\lambda} [\lVert \textbf{y}-\textbf{D}\mu_{\textbf{x}} \rVert_{2}^{2} + \hat{\lambda} \mathbb{E}_{q(\boldsymbol{\alpha};\tilde{\boldsymbol{\theta}}^{OLD})} \text{Tr}(\Sigma_{\textbf{x}}(\Sigma_{\textbf{x}}^{-1} -\Sigma_{0}^{-1}))] \label{eq:23} \\
& = -\frac{ML}{2}\ln{\lambda} -\frac{1}{2\lambda} [\lVert \textbf{y}-\textbf{D}\mu_{\textbf{x}} \rVert_{2}^{2} + \hat{\lambda} [NL - \mathbb{E}_{q(\boldsymbol{\alpha};\tilde{\boldsymbol{\theta}}^{OLD})} \text{Tr}(\Sigma_{\textbf{x}}(\Sigma_{0}^{-1}))] \nonumber \\
& = -\frac{ML}{2}\ln{\lambda} -\frac{1}{2\lambda} [\lVert \textbf{y}-\textbf{D}\mu_{\textbf{x}} \rVert_{2}^{2} +  \hat{\lambda} [NL - \text{Tr}(\Sigma_{\textbf{x}}\mathbb{E}_{q(\boldsymbol{\alpha};\tilde{\boldsymbol{\theta}}^{OLD})}(\Sigma_{0}^{-1}))], \label{eq:24}
\end{align}
where Eq.~\eqref{eq:23} follows Eq.~\eqref{eq:18}, and $\hat{\lambda}$ denotes the estimated $\lambda$ in the previous iteration. By setting the derivative of Eq.~\eqref{eq:24} over $\lambda$ to zero, $\lambda$ can be estimated as follows: 
\begin{equation*}
    \lambda=\frac{[\lVert \textbf{y}-\textbf{D} \mu_{\textbf{x}} \rVert_{2}^{2} + \hat{\lambda} [NL - \text{Tr}(\Sigma_{\textbf{x}}\mathbb{E}_{q(\boldsymbol{\alpha};\tilde{\boldsymbol{\theta}}^{OLD})}(\Sigma_{0}^{-1}))]]}{ML}.
\end{equation*}

\subsection{\emph{Fault pattern matrix \texorpdfstring{$\Phi$}{TEXT} in Section~\ref{s:sec5}}}\label{app:app6}
\begin{figure}[!htbp]
    \centering
    \resizebox{\textwidth}{!}{\includegraphics{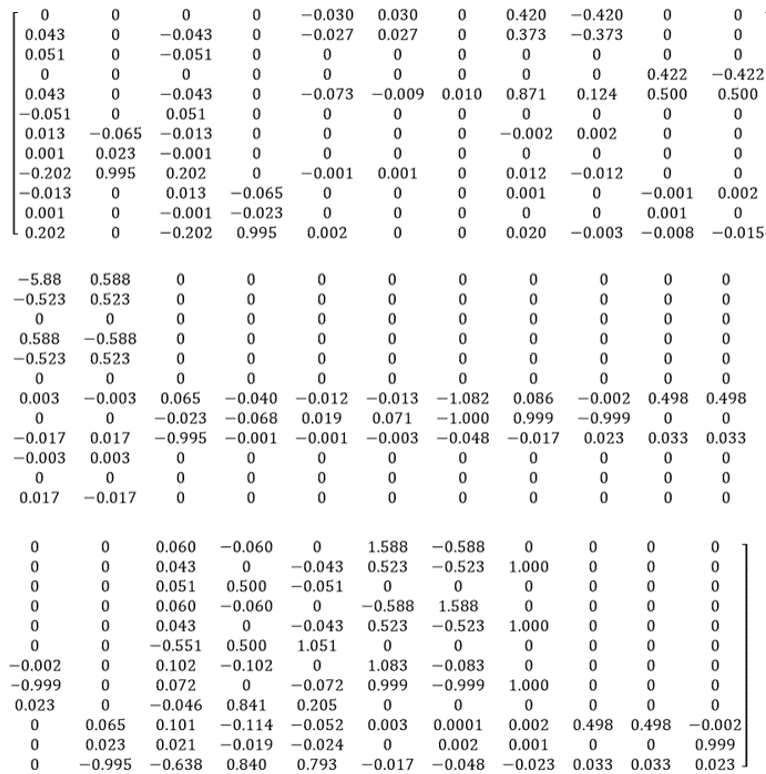}}\vspace{-0.5cm}
    \caption{Fault pattern matrix $\Phi$ in Section~\ref{s:sec5} \citep{bastani2018fault}.}
    \label{fig:fig11}
\end{figure}

Figure~\ref{fig:fig11} shows the fault pattern matrix $\Phi$ from previous researches \citep{bastani2016compressive, huang2007stream1, kong2009variation} that used in Section~\ref{s:sec5}.

\end{document}